\documentclass[aps,prd,twocolumn,superscriptaddress,groupedaddress,preprintnumbers]{revtex4}

\usepackage[english]{babel}
\usepackage{array}
\usepackage{tabulary}
\usepackage{amsfonts}
\usepackage{amssymb,amsmath}
\usepackage{cancel}
\usepackage{mathcomp}
\usepackage{graphicx}
\usepackage{color}
\usepackage{comment}
\usepackage{float}
\usepackage{placeins}
\usepackage[export]{adjustbox}

\def\beq{\begin{equation}}
\def\eeq{\end{equation}}
\def\baq{\begin{align}}
\def\eaq{\end{align}}
\def\d{{\rm d}}

\begin{document}

\title{Adiabatic perturbations in coupled scalar field cosmologies}

\preprint{}

\author{J.~Beyer$^a$ \\
{\it $^a$Institut f\"ur Theoretische Physik,   
Universit\"at Heidelberg,   
Philosophenweg 16, 69120 Heidelberg, Germany}}

\begin{abstract}
\begin{center}

We present a comprehensive and gauge invariant treatment of perturbations around cosmological scaling solutions for two canonical scalar fields coupled through a common potential in the early universe, in the presence of neutrinos, photons and baryons, but excluding cold dark matter. This setup is relevant for analyzing cosmic perturbations in scalar field models of dark matter with a coupling to a quintessence field. We put strong restrictions on the shape of the common potential and adopt a matrix-eigensystem approach to determine the dominant perturbations modes in such models. Similar to recent results in scenarios where standard cold dark matter couples to quintessence, we show that the stability of the adiabatic perturbation mode can be an issue for this class of scalar field dark matter models, but only for specific choices of the common potential. For an exponential coupling potential, a rather common shape arising naturally in many instances, this problem can be avoided. We explicitly calculate the dominant perturbation modes in such scenarios.

\end{center}
\end{abstract}

\maketitle

\section{Introduction}
\label{sec:Intro}
Scalar fields can appear in modern cosmology in many forms. They are typically associated with phenomena which can not be attributed to ordinary forms of matter, most commonly with an accelerated expansion of the universe. Thus they are particularly important in the very early universe for modeling the era of inflation \cite{Sato:1981ds,Guth:1981uk,Guth:1980zm,Linde:1981mu,Albrecht:1982wi,Lyth:1998xn}, and the very late universe, where scalar quintessence models are one of the most promising candidates for explaining dark energy \cite{Wetterich:1987fk,Wetterich:1987fm,Ratra:1987rm,Peebles:1987ek,Wetterich:1994bg,Peebles:2002gy,Copeland:2006wr}.

For the long time between these eras, scalar fields typically play only a supporting role in the cosmic evolution. From reheating onwards, throughout the entire radiation dominated era, their energy densities only contribute a small percentage to the total energy density of the universe \cite{Bean:2001wt,Calabrese:2011hg,Reichardt:2011fv,Xia:2013dea,Pettorino:2013ia}. The same holds true for the matter-dominated era up to redshifts of about 5, with the exception of scalar field models of dark matter \cite{UrenaLopez:2000aj,Matos:1999et,Matos:2000ng,Hu:2000ke,Matos:2000ss,Alcubierre:2001ea,Beyer:2010mt}. Still, much effort has ben put into analyzing scalar field dynamics in these eras. The reason for this lies in the so called coincidence problem, which adresses the question why the energy densities of dark energy and matter are roughly of the same order of magnitude only in the recent past. In the $\Lambda$CDM model this seems like a very precise finetuning of the cosmological constant. Scalar field models of dark energy provide a possible solution to this problem, since they can exhibit a behavior often called tracking or scaling \cite{Copeland:1997et,Liddle:1998xm,Steinhardt:1999nw,Zlatev:1998tr}, where the field equations have solutions for which the dynamics follow a specific trajectory largely irrespective of the initial conditions chosen, thereby explaining why dark energy became dominant only recently. 

In this paper we are interested in such scaling solutions involving multiple scalar fields which couple through their common potential and the evolution of perturbations in such scenarios. Similar models have been investigated before in the context of inflation \cite{Gordon:2000hv,Malik:2002jb}, i.e. without any additional matter content. We consider the dynamics of multiple scalar fields in the presence of ordinary matter, i.e baryons, neutrinos and photons, but exclude cold dark matter or a cosmological constant. This setting is relevant for models where the entire dark sector, i.e. both dark energy and dark matter is described by scalar fields. 

If one looks beyond the homogeneous and isotropic background cosmology, one quickly finds that the the inhomogeneities of scalar fields in such models are of great importance for the study of cosmological perturbations in general. This starts with the fact that all structures seen in the universe today are now widely believed to have been sourced by quantum fluctuations of scalar fields during the era of inflation, but does not end there. Scalar quintessence models can influence late-time structure growth \cite{Bartelmann:2005fc,Doran:2001rw}, but even during the era of radiation domination, where scalar fields are subdominant, their presence can impact how perturbations in the baryonic or dark matter sector evolve. While standard minimally coupled one-field scaling quintessence models seem to be unproblematic in this respect \cite{Doran:2003xq}, recent studies have shown that a coupling between a scalar field and dark matter can result in an instability of the adiabatic perturbation mode \cite{Majerotto:2009zz,Majerotto:2009np}, potentially rendering such scenarios difficult to reconcile with observational bounds, which clearly demand largely adiabatic perturbations \cite{Enqvist:2000hp,Enqvist:2001fu,Beltran:2005xd,Seljak:2006bg,Keskitalo:2006qv,Kawasaki:2007mb,Castro:2009ej,Valiviita:2009bp,Li:2010yb,Ade:2013uln}. It is therefore interesting to investigate whether such problems can also arise in models where multiple scalar fields couple to each other, e.g. in coupled scalar field dark matter models, and we do so in our analysis below.

This paper is organized as follows: 

In section \ref{sec:GenericScalingSolutions} we derive the generic shape of the potential necessary for multiple canonical scalar fields to exhibit scaling solutions in the presence of a background fluid. In section \ref{sec:EarlyUniversePerturbations} we study the evolution of linear perturbations in the superhorizon regime in such models, in the presence of photons, baryons and neutrinos. We restrict ourselves to the case of two scalar fields here for simplicity. In section \ref{sec:coupledExponentials}, we apply the results to one particular choice of scalar potential with an exponential coupling. This is the relevant potential for the coupled scalar dark matter model investigated in an accompanying paper \cite{CB_LinPers}. Finally, in section \ref{app:otherModels} we investigate alternative potentials and construct an explicit example for which the adiabatic perturbation mode is unstable. We present our conclusions in section \ref{sec:Conclusion}.

\section{Coupled canonical scalar fields and scaling solutions}
\label{sec:GenericScalingSolutions}

In this section we investigate exact scaling solutions in FLRW-cosmologies involving multiple coupled canonical scalar fields and a background matter fluid, defined by an energy density $\rho_m$ and a constant equation of state $\omega_m$. In the literature the definition of a scaling solution is not completely unambiguous. In the case of a single scalar field it can either mean a solution for which the scalar energy density scales like the scale factor to some constant power, or, and this can be considered a stricter definition, a solution for which the scalar energy density scales exactly like the background fluid. In this work we adopt the stricter definition and consider an exact scaling solution to be a scenario in which all scalar energy densities scale like the background fluid.

It is well known that the only potential providing an exact scaling solution for a single canonical scalar field is an exponential potential. This even holds true in the presence of a coupling to the background fluid \cite{Amendola:1999qq,Uzan:1999ch,Copeland:2004qe,Tsujikawa:2004dp}. Furthermore, even in the presence of non-canonical kinetic terms - however still restricted to those yielding second order field equations - the form of possible Lagrangians yielding scaling solutions can be strongly restricted \cite{Tsujikawa:2004dp}. 

We will now show that in the case of multiple canonical coupled scalar fields similar restrictions for the common potential can be found. To do this we employ an approach already used in the earlier papers cited above, slightly adjusted to fit our scenario. We start by considering the following action
\beq
S = \int \d^4x \sqrt{-g} \left[ \frac{R}{2} + \sum_{i=1}^n X_{\varphi_i} + V(\varphi_1,..., \varphi_n) \right] + S_{\rm m} \, ,
\eeq
where $X_{\varphi_i}=\frac{1}{2} \mathcal{D}_\mu \varphi_i \mathcal{D}^\mu \varphi_i$, $n$ denotes the number of scalar fields $\varphi_i$ present and $S_{\rm m}$ denotes the matter action. The scalar field equations for this action in the FLRW-background read
\beq
\label{scalarFEQ}
\varphi_i'' + 2 h \varphi_i' +a^2 V_{,\varphi_i} = 0 \, ,
\eeq
where a prime denotes a derivative with respect to conformal time and $h$ is the conformal hubble parameter $h=a'/a$.
The common potential $V$ can in principal be identified with any one of the scalar field energy densities or even split in some arbitrary fashion, but we choose to assign it to $\varphi_1$ for simplicity and define scalar energy densities and pressure densities as follows:
\begin{align}
\label{potSplit1}
\rho_{\varphi_1}= X_{\varphi_1} + V\, , \quad p_{\varphi_1} = X_{\varphi_1} - V \, , \\
\rho_{\varphi_j} = p_{\varphi_j} = X_{\varphi_j} \, , \quad j = 2,...,n \, .
\end{align}
As we assume no direct coupling between the background fluid and the scalar sector, the equations governing the background-evolution can be written as
\begin{align}
\label{matterEEC}
&\frac{\d \rho_m}{\d N} + 3 (1+\omega_m) \rho_m = 0 \, , \\
\label{varphiEEC}
&\frac{\d \rho_{\varphi_i}}{\d N} + 3 (1+\omega_{\varphi_i}) \rho_{\varphi_i} = 3 (1+\omega_{\varphi_i}) q_i \rho_{\varphi_i} \, ,
\end{align}
where $N={\rm ln}(a)$ and conservation of the total energy-momentum tensor implies that the scalar couplings have to satisfy 
\beq
\label{totalCouplingZero}
\sum_i (1+\omega_{\varphi_i}) \rho_{\varphi_i} q_i = 0 \, .
\eeq
The definition of the couplings $q_i$ in this manner is of course a matter of convention, here we remain consistent with the general formulas given in appendix \ref{app:LinearPerturbations}.
In an exact scaling scenario we demand
\begin{align}
\label{scalingCondition}
&\omega_{\varphi_i} = \frac{p_{\varphi_i}}{\rho_{\varphi_i}}={\rm const.}  \quad {\rm and} \quad
\frac{\rho_{\varphi_i}}{\rho_m} = {\rm const.} \, .
\end{align}
Note that the scalar equations of state can differ from the background equation of state even in an exact scaling scenario if a coupling between the scalar fields is present, only the combined scalar equation of state has to fulfill
\beq
\label{omegaScalarEQ}
\omega_{\rm sc} = \frac{\sum_i \rho_{\varphi_i} \omega_{\varphi_i}}{\sum_i \rho_{\varphi_i}} =\omega_m \, ,
\eeq
if the scalar density parameters are non-zero. 
From equations (\ref{matterEEC}) and (\ref{scalingCondition}) we can directly conclude that
\beq
\label{rhoScaling}
\frac{\d {\rm ln} \rho_{\varphi_i}}{\d N} = \frac{\d {\rm ln} \rho_m}{\d N} = -3(1+\omega_m) \, ,
\eeq
and 
\beq
\label{xScaling}
\frac{\d {\rm ln} X_{\varphi_i}}{\d N} = -3(1+\omega_m) \, .
\eeq
Since $X_{\varphi_i} \propto \left(\d \varphi_i / \d N\right)^2 \rho_{\rm tot}$, equations (\ref{rhoScaling}) and (\ref{xScaling}) directly give 
\beq
\label{constDerivs}
\d \varphi_i / \d N =c_i = {\rm const.} \, .
\eeq
Furthermore all couplings have to be constant by virtue ofequations (\ref{varphiEEC}) and (\ref{rhoScaling}): 
\beq
\label{couplingConstEQ}
q_i =  \frac{(\omega_{\varphi_i}- \omega_m)}{1+\omega_{\varphi_i}} \, .
\eeq
Now we can easily derive a differential equation for the potential $V$ by again employing equations (\ref{varphiEEC}) and (\ref{rhoScaling}):
\beq
\frac{\d {\rm ln} V}{\d N} = \frac{\d {\rm ln} V}{\d {\varphi_i}} c_i  = -3(1+\omega_m),
\eeq
which directly gives
\beq
\label{lnVConstraint}
{\rm ln} \, V = -\frac{3(1+\omega_m)}{c_1} \varphi_1 + f \left(\xi_2,...,\xi_n \right) \, ,
\eeq
where $\xi_i = \varphi_i - \frac{c_i}{c_1} \varphi_1$ and $f$ is some arbitrary (smooth) function. One can quickly check that this solution is not in conflict with the the demand of constant couplings simply by noting that  by virtue of equation (\ref{constDerivs}) all $\xi_i$ have to be constant for a scaling solution and by convention (\ref{potSplit1}) we therefore have
\beq
q_j = \frac{-V_{,\varphi_j} c_j}{3 (1+\omega_{\varphi_j}) \rho_{\varphi_j}} = - \frac{(1-\omega_{\varphi_1}) c_j \rho_{\varphi_1}}{6(1+\omega_{\varphi_j}) \rho_{\varphi_j}} \frac{\partial f}{\partial \xi_j} = {\rm const.}  
\eeq
for $j>2$ and dynamics corresponding to a scaling solution.

One comment might be in order here: In the derivation of equation (\ref{lnVConstraint}) we made some implicit assumptions. The most obvious one is that $\omega_{\varphi_1} \neq 1$, i.e. the scalar potential energy density should be non-vanishing. As we will see in section \ref{sec:coupledExponentials}, purely kinetic fixed points with $\omega_{\varphi_1} = 1$ do exist, and we expect them to be present for a large class of potentials, including many not of the shape derived here. The only restriction here is that the common potential should become zero somewhere, at least asymptotically. Furthermore one might suggest that we have excluded solutions for which cosmological expansion is purely scalar field dominated. This is however not the case, as one can simply replace all $\omega_m$ appearing in the above equations with a constant effective equation of state $\omega_{\rm eff}$ and reach the same conclusions concerning the potential shape.

To clarify this result: If you want to construct a potential allowing for exact scaling solution with non-vanishing potential energy density involving $n$ multiple canonical scalar fields coupled through their potential, you first have to check if your potential fulfills equation (\ref{lnVConstraint}) for some set of constants $c_1, ... , c_n$. If so, a scaling solution can exist.

\section{Early Universe Perturbations}
\label{sec:EarlyUniversePerturbations}

We will now turn our attention to the evolution of perturbations in the early universe in models containing multiple non-minimally coupled canonical scalar fields. In order to keep things simple, we restrict our attention to the case of two fields and call them $\varphi$ and $\chi$, deviating from the conventions used in section \ref{sec:GenericScalingSolutions}. We also split the potential differently, this time assigning a $\varphi$-dependent part to $\rho_\varphi$ and a common part to $\rho_\chi$, i.e.
\begin{align}
\label{potSplitMain}
&V=V_1 (\varphi) + V_2(\varphi,\chi) \, , \\
&\rho_\varphi = X_\varphi + V_1 \, , \quad \rho_\chi = X_\chi + V_2 \, .
\end{align}

By "early universe" we mean an era during which the cosmic expansion proceeds as if dominated by radiation, i.e. $h(\eta) =\eta^{-1}$, with $\eta$ being conformal time. We will consider a background-evolution during which both fields follow an exact scaling bahaviour, in particular the equations of state are assumed to be constant. Note that this does not imply a strongly subdominant role of the scalar fields or equations of state of exactly $1/3$. The coupling between the two fields enables a wide range of possible $\omega_\varphi$'s and $\omega_\chi$'s while still allowing for a radiation-like expansion ($a(\eta) \propto \eta$) and non-negligible scalar field contributions to the energy-densities already in very simple models, as we will see below.

In addition to the two coupled scalar fields we include neutrinos, photons and baryons into our model, but exclude a possible cold dark matter contribution. This sets our study apart from former systematic studies of early universe perturbation modes, which  have usually been limited to cosmologies containing either no scalar fields \cite{Ma:1995ey}, only scalar fields (usually in the context of inflation) \cite{Gordon:2000hv} or a single quintessence field \cite{Malik:2002jb,Malquarti:2002iu,Bartolo:2003ad,Doran:2003xq,Majerotto:2009np}. Leaving out a possible cold dark matter contribution is realistic, if at least one of the two scalar fields changes its dynamics away from a scaling solution and acts like a dark matter component during the later stages of its evolution. If this is the case, the coupling arising from the common potential will result in a coupled quintessence model in the late universe 
\cite{Wetterich:1994bg,Amendola:1999er,Pourtsidou:2013nha}, more specifically a coupled quintessence-scalar-dark-matter model \cite{Beyer:2010mt}. It is precisely these kinds of models that require the analysis presented here to draw the initial conditions for numerical evolutions of cosmological perturbations, a study we performed in an accompanying work \cite{CB_LinPers}.

Before moving on to our analysis, let us recap some relevant results from the previous studies. As was shown in \cite{Malik:2002jb}, even in generic non-minimally coupled models (which of course includes ours), purely adiabatic perturbations remain purely adiabatic on superhorizon scales, no matter what the interaction is. Here a mode is called adiabatic if all entropy perturbations vanish. This includes relative entropy perturbations between different components of the cosmic fluid as well as internal entropy perturbations for  a single component - these can exist for non-perfect fluids (see section 4 in appendix \ref{app:LinearPerturbations} for more details). However, as was noted in ref. \cite{Doran:2003xq}, the existence of such an adiabatic mode is non-trivial, since demanding all entropy perturbations to vanish typically introduces more constraints than we have degrees of freedom in the equations, in particular if the number of species considered is large. One should understand that this is not in conflict with the theorem found by Weinberg in \cite{Weinberg:2003sw,Weinberg:2003ur}, which states that an adiabatic superhorizon mode always exists, no matter the content of the universe. The apparent contradiction stems from Weinberg's definition of an adiabatic superhorizon-mode, which he takes to mean a mode for which the total curvature perturbation $\zeta$ remains constant, which is the case if the total entropy perturbation $\Gamma_{\rm tot} = 0$ (see appendix \ref{app:LinearPerturbations}). This is a weaker condition than demanding all entropy perturbations to vanish, which is what we take adiabatic to mean in this work. 

Furthermore this result is not sufficient to explain a suppression non-adiabatic modes, since any, presumably initially small, admixture of such modes could in principle eventually outgrow the adiabatic perturbations. Such a growth can happen for example during the 'adjustment' phases in minimally coupled tracking quintessence models when the quintessence field does not yet follow the tracker trajectory. However, a sufficiently long tracking regime typically erases these modes in this case \cite{Malquarti:2002iu}. 

The situation in non-minimally coupled models of quintessence appears to be more complicated. We are aware of only one study which systematically analyzes the evolution of superhorizon-perturbations in a coupled quintessence model \cite{Majerotto:2009np}. It appears that in such scenarios there are certain sections of parameter space for which the adiabatic mode becomes unstable, i.e. other, faster growing perturbation modes exist. Since this analysis includes a conventional cold dark matter component and uses a specific commonly used form of the coupling given by $Q_\varphi = - \beta \rho_{\rm cdm}$, the question if similar issues will arise in coupled scalar field models of dark matter remains open, and we will address it below. 

\subsection{Basic Formalism}

The basic idea we employ in our study of early universe superhorizon perturbations is one already used in earlier works \cite{Malquarti:2002iu,Bartolo:2003ad,Doran:2003xq}. We write the differential equations governing the evolution of linear perturbations in a convenient matrix form, i.e. 
\beq
\label{MatrixEquation}
\frac{\d U(x)}{\d \, {\rm ln}(x)} = A(x) U(x) \, .
\eeq
Here we introduced a convenient new time variable $x \equiv k/h$ and combined all relevant perturbative quantities into a single perturbation vector $U(x)$. Which quantities these are depends on the approximations one choses to use in the neutrino-, photon- and baryonic sectors. Here we employ the simplest (lowest order) version of the tight-coupling approximation for photons and baryons, which yields a common velocity potential for both components and excludes all higher momenta of the Boltzmann hierarchy for photons. In the case of neutrinos we truncate the Boltzmann-expansion after the quadrupole, leaving an additional anisotropic stress contribution. More details and precise definitions of the perturbative quantities used in this chapter can be found in appendix \ref{app:LinearPerturbations}. 

Since the scalar fields only have two degrees of freedom, the components of the perturbation vector can be chosen to be the energy density contrasts $\Delta_\alpha$ and velocity potentials $V_\alpha$ for all components of the cosmic fluid as well as anisotropic stress for neutrinos, i.e. 
\beq
U = \{\Delta_\nu, V_\nu, \Delta_\gamma, \Delta_b, V_{\gamma b} , \Delta_\varphi, V_\varphi,\Delta_\chi,V_\chi, \tilde{\Pi}_\nu  \} \, .
\eeq
The entries of the perturbation matrix $A$ can now be read off from the differential equations governing the evolution of these variables. For the photon-, baryon- and neutrino-sector they read
\begin{align}
\label{dDeltagammaMain}
\frac{\d \Delta_\gamma}{\d {\rm ln}(x)} = & - \frac{4}{3} x^2 V_{\gamma b}  \, , \\
\frac{\d \Delta_\nu}{\d {\rm ln}(x)} = & - \frac{4}{3} x^2 V_\nu \, , \\
\frac{\d \Delta_b}{\d {\rm ln}(x)} = & -x^2 V_{\gamma b} \, , 
\end{align}
\begin{align}
\frac{ \d V_\nu}{\d {\rm ln}(x)} = & \frac{1}{4} \Delta_\nu - V_\nu + 2 \Psi - \left( \frac{1}{6}  x^2 + \Omega_\nu \right) \tilde{\Pi}_\nu \\
\label{dVgammabMain}
\frac{ \d V_{\gamma b}}{\d {\rm ln}(x)} = & \frac{4 \Omega_\gamma}{4 \Omega_\gamma + 3 \Omega_b} \left(\frac{1}{4} \Delta_\gamma + \Psi \right) - \frac{4 \Omega_\gamma + 6 \Omega_b}{4 \Omega_\gamma+3 \Omega_b} V_\gamma + \Phi \, , \\
\label{dPinuEquationMain}
 \frac{\d \tilde{\Pi}_\nu}{\d {\rm ln}(x)}  = &  \frac{8}{5} V_\nu - 2 \tilde{\Pi}_\nu \, .
\end{align}
These equations can be derived from the generic equations given in appendix \ref{app:LinearPerturbations} by setting $\omega_{\rm eff}=1/3$. Note that the equation for the velocity potential $V_{\gamma b}$ is different from the one usually employed, e.g. in refs. \cite{Doran:2003xq,Majerotto:2009np}. This version is the correct one if one makes no assumption about the ratio $\Omega_b/\Omega_\gamma$. Both version agree in the limit $\Omega_b/\Omega_\gamma \rightarrow 0$.

The gravitational potentials are then given by
\begin{align}
\Psi =& -\frac{3}{2} \frac{\sum_\alpha \Omega_\alpha \left( \Delta_\alpha + 3 (1+\omega_\alpha) V_\alpha \right)}{x^2 + 6} \, , \\
\Psi'/h =& -\Phi + \frac{3}{2} \sum_\alpha \Omega_\alpha (1+\omega_\alpha) V_\alpha \, , \\
\Phi =& \Psi - 3 \sum_\alpha \Omega_\alpha \omega_\alpha \tilde{\Pi}_\alpha = \Psi - \Omega_\nu \tilde{\Pi}_\nu\, .
\end{align}

The generic equations of motion for the scalar fields are quite complicated (see appendix \ref{app:LinearPerturbations}), but they can be simplified considerably in an exact scaling scenario. The background quantities appearing in the scalar sector are the equations of state $\omega_\varphi$ and $\omega_\chi$, the adiabatic sound speeds $c_{a,\varphi}^2$ and $c_{a,\chi}^2$, the couplings $q_\varphi$ and $q_\chi$, their time-derivatives and the second derivative of the common potential $V_{2,\varphi \varphi}$.

In an exact scaling scenario we assume constant equations of state, i.e. we set $\omega_\varphi'=\omega_\chi'=0$. This implies that the adiabatic sound speeds are given by $c_{a,\varphi}^2=\omega_\varphi$ and $c_{a,\chi}^2=\omega_\chi$. Furthermore the couplings $q_\varphi$ and $q_\chi$ have to be constant by equation (\ref{couplingConstEQ}) and are of course related by equation (\ref{totalCouplingZero}). Finally the scalar equations of state are related by equation (\ref{omegaScalarEQ}) and we only have to deal with the second derivative $V_{2,\varphi \varphi}$ of the common potential. For a generic scalar field evolution this can of course take any, in principle time-dependent, form, but in a scaling scenario it merely introduces another constant into the equations, which can be seen as follows.

We start by investigating the generic potential for exact scaling solutions given by equation (\ref{lnVConstraint}). In order to split our potential according to equation (\ref{potSplitMain}) we have to rewrite $f$ as
\beq
\label{ftog}
f(\xi) = {\rm ln} \left( \mu_1 + \mu_2 g(\xi) \right) \, ,
\eeq
where $\xi = \chi - c_2/c_1 \varphi$ and $\mu_1$, $\mu_2$ are some suitable constants. Note that this already shows that $V_1$ has to be an exponential potential. We can take the second derivative of this potential with respect to $\varphi$ and obtain (for $\omega_m=1/3$)
\beq
V_{2,\varphi \varphi} = \mu_2 {\rm e}^{-4 \varphi/c_1} \left[ \frac{16}{c_1^2} g(\xi) + \frac{8 c_2}{c_1^2} g'(\xi) + \frac{c_2^2}{c_1^2} g''(\xi) \right] \, .
\eeq
Since $\xi$ is constant during a scaling solution, it is obvious that $V_{2,\varphi \varphi} \propto V \propto \rho_{\rm tot}$ in such scenarios and we can thus define
\beq
V_{2,\varphi \varphi} = 3 q_\varphi \frac{h^2}{a^2} r_\varphi \, ,
\eeq
with some constant $r_\varphi$. Note that $r_\varphi$ is in principle a new independent constant depending on the functional form of $g$ and can generally generally not be related to $q_\varphi$, as it depends on the second derivative $g''$, whereas only the first derivative $g'$ enters into $q_\varphi$.

Now we can finally write down the equation for the scalar sector in a form where the only remaining constant parameters are $\omega_\varphi$, $r_\varphi$ and the density parameters. They read
\begin{align}
\label{dDeltaVarphiMain}
\frac{\d \Delta_\varphi}{\d {\rm ln}(x)} = & 
\frac{\omega_\varphi - 3}{1+\omega_\varphi} \Delta_\varphi 
+ (3 \omega_\varphi - 1) \left( \Psi'/h + \Phi \right)  \nonumber \\
& + (3 \omega_\varphi - 1) (r_\varphi-4) V_\chi + \frac{4 (\omega_\varphi - 3)}{1+\omega_\varphi} \Psi\nonumber \\
& + \Big( (1-3\omega_\varphi) r_\varphi -x^2 (1+\omega_\varphi)  \nonumber \\
& \hspace{15pt} + \frac{2 (\omega_\varphi - 1)(7+3\omega_\varphi)}{1+\omega_\varphi} \Big) V_\varphi \, , \\
\frac{\d V_\varphi}{\d {\rm ln}(x)} =& \frac{1}{1+\omega_\varphi} \Delta \varphi + \Phi +\frac{2(1- \omega_\varphi)}{1 + \omega_\varphi} V_\varphi+\frac{4}{1 + \omega_\varphi} \Psi \, , \\
\frac{\d \Delta_\chi}{\d {\rm ln}(x)} = & -2 \Delta_\chi + \frac{\Omega_\varphi}{\Omega_\chi} (1-3\omega_\varphi) \left( \Psi'/h + \Phi \right) \nonumber \\
& - 4 \left( 2 + \frac{\Omega_\varphi}{\Omega_\chi} \frac{1-3\omega_\varphi}{1+\omega_\varphi} \right) \Psi + \frac{\Omega_\varphi}{\Omega_\chi} \frac{1-3\omega_\varphi}{1+\omega_\varphi} \Delta_\varphi \nonumber  \\
& + \frac{\Omega_\varphi}{\Omega_\chi} (3 \omega_\varphi -1) \left(4 \frac{ 2+ \omega_\varphi}{1+\omega_\varphi} -r_\varphi \right) V_\varphi \nonumber \\
& - \Bigg( \frac{(r_\varphi - 6) (3\omega_\varphi-1)\Omega_\varphi + 8 \Omega_\chi}{\Omega_\chi} \nonumber \\
& \hspace{15pt} + \frac{4 \Omega_\chi + \Omega_\varphi (1-3\omega_\varphi)}{3 \Omega_\chi} x^2 \Bigg) V_\chi  \, , \\
\label{dVChiMain}
\frac{\d V_\chi}{\d {\rm ln}(x)} =& \frac{3(1-3 \omega_\varphi ) \Omega_\varphi }{(3 \omega_\varphi-1) \Omega_\varphi - 4 \Omega_\chi } V_\varphi + \frac{12 \Omega_\chi }{\Omega_\varphi (1-3 \omega_\varphi) +4 \Omega_\chi } \Psi \nonumber \\
& + \frac{3 \Omega_\chi }{\Omega_\varphi (1-3\omega_\varphi) +4 \Omega_\chi } \Delta_\chi  +\Phi + V_\chi \, .
\end{align}
The coefficients of the matrix $A(x)$ can be read off from equations, (\ref{dDeltagammaMain}) -  (\ref{dPinuEquationMain}) and (\ref{dDeltaVarphiMain}) - (\ref{dVChiMain}) after replacing the gravitational potentials. 

Before analyzing the solutions for the perturbations we have to investigate the background quantities appearing in $A(x)$. Those are generally $x$-dependent, but in this analysis we are interested only in the superhorizon-limit defined by $x \ll 1$ and we can therefore work with Taylor-expansions in $x$. To see that, first note that we assume a radiation-like expansion with a scalar scaling solution in the early universe, a scenario only slightly disturbed by the baryonic density parameter, which is small, but grows linearly in $x$:
\beq
\Omega_b = \Omega_{b,{\rm in}} \frac{a}{a_{\rm in}} = \Omega_{b,{\rm in}} \frac{h_{\rm in}}{k} x \, .
\eeq 
This implies that all the other density parameters are constant to leading order, but decrease slowly at linear order in $x$ . Evaluating Friedmanns equations for a flat universe in the form $\Omega_\nu + \Omega_\gamma + \Omega_\varphi + \Omega_\chi + \Omega_b = 1$ order by order in $x$ then yields to leading order 
\beq
\Omega_\alpha = \Omega_{\alpha,0} \left( 1-x \right) \quad \text{for all $\alpha \neq b$.}
\eeq
Using this, it becomes obvious that we can expand the matrix $A(x)$ as a type of Taylor-series in x and recover a constant matrix at leading order:
\begin{align}
A(x) =  A_0 & + A_1x + A_2x^2 
+ \mathcal{O}(x^3) \, .
\end{align}
With this approximation we can immediately write down the leading order solution for equation (\ref{MatrixEquation}), which reads
\beq
U_0(x) = \sum_i c_i x^{\lambda_i} U_0^{(i)} \, ,
\eeq
where $U_0^{(i)}$ are eigenvectors of the matrix $A_0$ and $\lambda_i$ the corresponding eigenvalues. 
With this as a starting point, one can now easily determine higher order corrections to the found solution by expanding the eigenvectors as follows:
\begin{align}
U^{(i)}(x) = U_0^{(i)} &+ U_1^{(i)} x + U_2^{(i)} x^2 
+ \mathcal{O}(x^3) \, ,
\end{align}
where the first order correction is then given by
\beq
U_1^{(i)} = \left((\lambda_i+1) \mathbb{I}-A_0\right)^{-1} A_1 U_0^{(i)} \, ,
\eeq
and the other corrections can be calculated in a similar fashion, but we will not need them here. Of course, the general solution is given by
\beq
U(x) = \sum_i c_i x^{\lambda_i} U^{(i)}(x) \, .
\eeq
A short comment concerning the different approximations used here might be in order. As explained in appendix \ref{app:LinearPerturbations} we use the simplest version of the tight coupling approximation for photons and baryons, i.e. we only include coupling terms to leading order (that is zeroth order) of $h \tau_c \ll 1$, where $\tau_c = (a n_e \sigma_T)^{-1}$ is the Thomson interaction timescale. One might question whether it is consistent to go to next to leading order in the $x$-expansion of the matrix $A(x)$, but not include the next order in the TCA, i.e. terms suppressed by $h \tau_c$. As it turns out, for the relevant wavenumbers we have $x=k/h \gg h \tau_c$ during radiation domination. This breaks down for very small $k$, but for realistic cosmologies the boundary lies at roughly $10^{-4}$ Mpc$^{-1}$, which is already far below what is currently observable.

\subsection{Dominant perturbation modes}

\subsubsection{Eigenvalues}

In order to determine the dominant perturbation modes in the early universe in coupled two scalar field models one simply has to determine the eigenvalues $\lambda_i$ of the matrix $A_0$ with the largest real parts and the corresponding eigenvectors. In principle these depend on the $4$ parameters determining the scaling solution, which can be taken to be $\omega_\varphi$, $r_\varphi$, $\Omega_\varphi$ and $\Omega_\chi$. The remaining density parameters for neutrinos and photons are fixed by the fact that we assume a FLRW metric without curvature, i.e. $\Omega_\nu + \Omega_\gamma + \Omega_\varphi + \Omega_\chi = 1$ and the well known relation $\Omega_\nu = 21/8 \times \left( 4/11 \right)^{4/3} \Omega_\gamma$, which is valid after electron-positron annihilation \cite{Bartelmann:2009te}. However, four of the eigenvalues are always given by $\left\{\lambda_1,\lambda_2,\lambda_3,\lambda_4 \right\} = \left\{0,0,0,-1\right\}$, irrespective of the values of these parameters. The remaining 6 eigenvalues are very complicated functions of the 4 free parameters and quoting them is not very enlightening. We will however present numerical results below, obtained simply by running through a grid of the four independent parameters $\{ \omega_\varphi,r_\varphi,\Omega_\varphi,\Omega_\chi \}$. 

\begin{figure}[t]
	\centering
  \includegraphics[width=0.75\linewidth,valign=t]{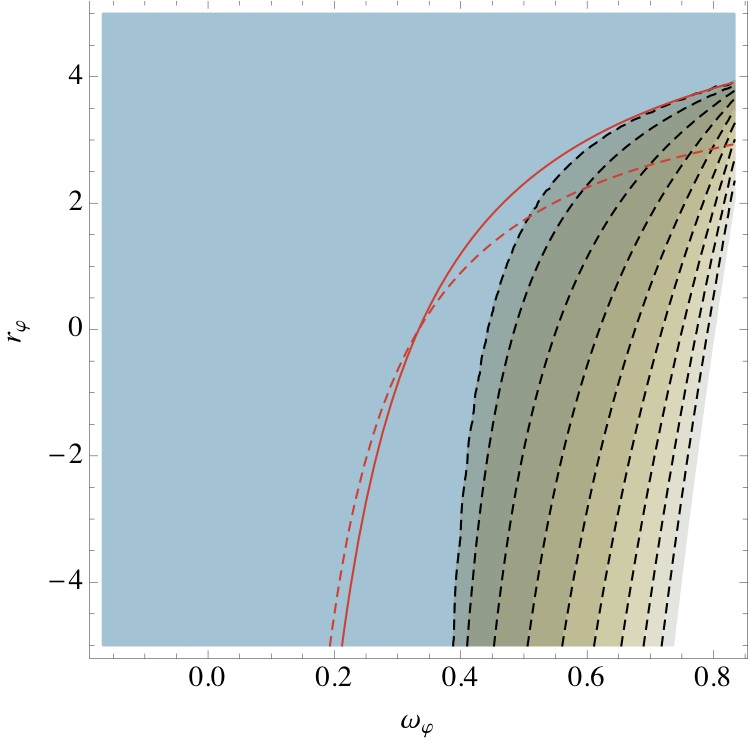}
  \hspace{5pt}
  \raisebox{-.3\height}{\includegraphics[width=0.08\linewidth,valign=t]{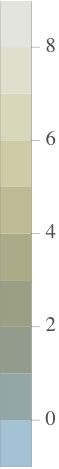}}
	\caption{Real part of one critical eigenvalue for $\Omega_\varphi=0.05$ and $\Omega_\chi=0.02$. The solid red line represents a model with an exponential potential discussed in section \ref{sec:coupledExponentials}, the dashed red line a power-law potential discussed in section \ref{app:otherModels}.}
	\label{fig:genericEV5_1}
\end{figure}
\begin{figure}[t]
	\centering
  \includegraphics[width=0.75\linewidth,valign=t]{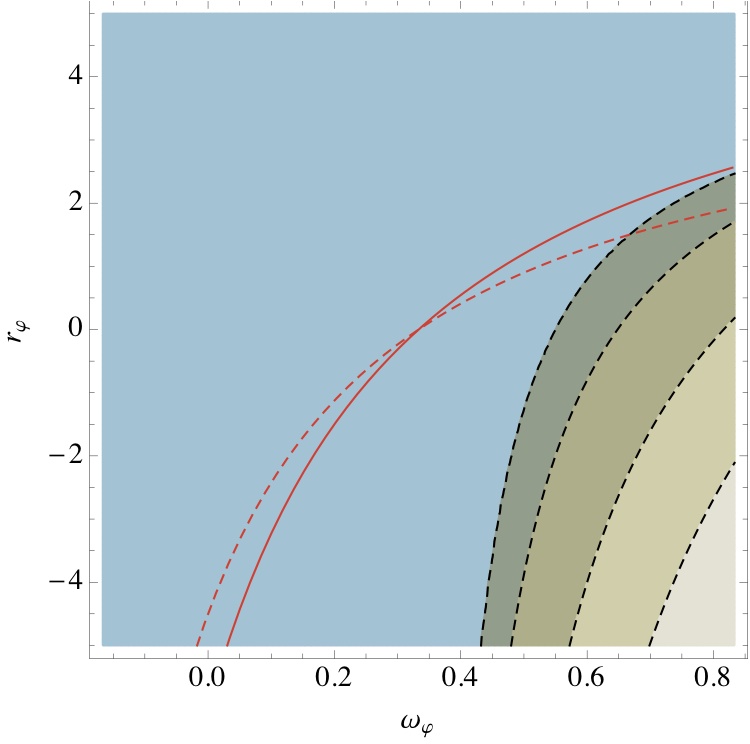}
  \hspace{5pt}
  \raisebox{-5.\height}{\includegraphics[width=0.08\linewidth,valign=t]{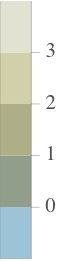}}
	\caption{Real part of one critical eigenvalue for $\Omega_\varphi=0.002$ and $\Omega_\chi=0.002$. The solid red line represents a model with an exponential potential discussed in section \ref{sec:coupledExponentials}, the dashed red line a power-law potential discussed in section \ref{app:otherModels}.}
	\label{fig:genericEV5_2}
\end{figure}

For the wide range of parameters we investigated, 4 of the 6 remaining eigenvalues had real parts which were bound from above by $-0.5$. Only two eigenvalues can have a real part bigger than zero, a case will will call \textit{strongly growing} from now on. As we will see below, the nullspace of $A_0$ always contains an adiabatic mode. It is precisely the two potentially strongly growing modes which can render the adiabatic mode unstable, which would make the corresponding scenario difficult to reconcile with observations \cite{Enqvist:2000hp,Enqvist:2001fu,Beltran:2005xd,Seljak:2006bg,Keskitalo:2006qv,Kawasaki:2007mb,Castro:2009ej,Valiviita:2009bp,Li:2010yb,Ade:2013uln}. 
On could of course conceivably come up with reasons why these modes are initially strongly suppressed compared to the adiabatic mode, so that it does not become dominant during the radiation dominated era, but such a fine-tuning of initial conditions is often not very natural and generally undesirable.  

To visualize the effect of the coupling between the two scalar fields we have plotted the real part of one of the two critical eigenvalues in Figures \ref{fig:genericEV5_1} and \ref{fig:genericEV5_2}. As can be seen, the values become bigger than zero for large deviations of $\omega_\varphi$ from the uncoupled value of $1/3$, i.e. for big couplings. Furthermore, as one would expect, this effect becomes stronger for larger scalar field density parameters. In the uncoupled case, even the critical eigenvalues have real parts bound by $-0.5$ from above and the corresponding modes are thus subdominant. This shows that a strong coupling can be relevant for early universe perturbations, because it can potentially change the dominant perturbation mode, rendering the adiabatic mode unstable.

One should note however, that just because we can find a set of parameters which yield a strongly growing eigenvalue in this numerical treatment does not necessarily mean that we can construct a model actually yielding a scaling solution with such a feature. The problem is of course, that for any given scaling solution some of the parameters treated here as free might well be related, or scaling solutions with a radiation like expansion might not even exist or be confined to specific regions of parameter-space. Later in this work we will examine one specific potential where the $r_\varphi-\omega_\varphi$ relation is fixed to the solid red line in Figures \ref{fig:genericEV5_1} and \ref{fig:genericEV5_2}, thus excluding the region of strongly growing eigenvalues, and a second one represented by the dashed red line, for which the adiabatic mode becomes unstable for strong couplings.

\subsubsection{Eigenvectors}

Since the generic eigenvalues have very complicated forms in general, so do the corresponding eigenvectors. This also holds for the eigenvectors belonging to the two critical eigenvalues, representing potentially dominant modes. Things look a lot better for the nullspace of $A_0$, which of course includes all dominant perturbations modes in the absence of strongly growing modes. One is free to choose any suitable basis of this three-dimensional space, we choose to categorize the modes according following ref. \cite{Doran:2003xq}, i.e. by means of the total curvature and relative entropy perturbations.

In this categorization a mode is called an isocurvature mode if the total curvature perturbation vanishes, i.e.
\beq
\zeta = \sum_\alpha \frac{\rho_\alpha \Delta_\alpha }{\rho_{\rm tot} + p_{\rm tot}}  = 0 \, ,
\eeq 
and adiabatic if all relative and internal entropy perturbations, i.e.
\begin{align}
S_{\alpha \beta} &= \frac{\Delta_\alpha}{(1+\omega_\alpha) (1-q_\alpha)} - \frac{\Delta_\beta}{(1+\omega_\beta) (1-q_\beta)} = 0 \, , \\
\Gamma_\alpha &= \frac{1}{p_\alpha} \left( \delta p_\alpha - c_{a,\alpha}^2 \delta \rho_\alpha \right) = 0 \quad \text{for all $\alpha$, $\beta$} \, .
\end{align}
There are some deviating definitions of the total curvature pertubation in the literature \cite{Mukhanov:1990me}, but they agree in the superhorizon limit in a flat universe.

This categorization splits the nullspace into a two-dimensional isocurvature space and one adiabatic mode. The adiabatic eigenvector is given by
\beq
\begin{pmatrix}
\Delta_\nu \\
V_\nu \\
\Delta_\gamma \\
\Delta_b \\
V_{\gamma b} \\
\Delta_\varphi \\
V_\varphi \\
\Delta_\chi \\
V_\chi \\
\Pi_\nu
\end{pmatrix}_{\rm adiab.}
= \mathcal{C}
\begin{pmatrix}
1 \\ -5/4 \, \mathcal{R}^{-1} \\ 1 \\ 3/4 \\-5/4 \, \mathcal{R}^{-1} \\ 1 \\-5/4 \, \mathcal{R}^{-1} \\1\\-5/4 \, \mathcal{R}^{-1} \\ - \mathcal{R}^{-1} 
\end{pmatrix} \, ,
\eeq
where $\mathcal{R}=(15 + 4\Omega_\nu)$. This is very similar to the adiabatic mode in minimally coupled quintessence scenarios \cite{Doran:2003xq}. 

The isocurvature-subspace can be further decomposed into one baryon-isocurvature and one neutrino/scalar-isocurvature mode.

The baryon isocurvature mode is characterized by the demand that all species except for baryons should be adiabatic, i.e. $S_{b,\alpha}\neq 0$ for all $\alpha$, but $S_{\alpha,\beta}=0$ for $\alpha, \beta \neq b$. For our set of scaling scenarios this necessarily implies that the scalar internal entropy perturbations also vanish, i.e. $\Gamma_\varphi = \Gamma_\chi = 0$. In the corresponding eigenvector all entries except for $\Delta_b$ vanish at leading order, we therefore quote the result to subleading order in $x$ here:
\beq
\begin{pmatrix}
\Delta_\nu \\
V_\nu \\
\Delta_\gamma \\
\Delta_b \\
V_{\gamma b} \\
\Delta_\varphi \\
V_\varphi \\
\Delta_\chi \\
V_\chi \\
\Pi_\nu
\end{pmatrix}_{\rm bar. iso.}
= \mathcal{C}
\begin{pmatrix}
0 \\ 
- 45/8 \, \Omega_b \mathcal{P}^{-1} / (15 +2 \Omega_\nu)  \\ 
0 \\ 
1 \\ 
- 45/8 \, \Omega_b \mathcal{P}^{-1} / (15 +2 \Omega_\nu)  \\ 
-3/4 \, (\omega_\varphi - 3) \Omega_b \mathcal{P}^{-1} \\ 
- 45/8 \, \Omega_b \mathcal{P}^{-1} / (15 +2 \Omega_\nu)  \\ 
-2 \Omega_b/\Omega_\chi \, \mathcal{P}^{-1} \mathcal{X}  \\
- 45/8 \, \Omega_b \mathcal{P}^{-1} / (15 +2 \Omega_\nu)  \\ 
- 3 \Omega_b \mathcal{P}^{-1} / (15 +2 \Omega_\nu)   
\end{pmatrix} \, ,
\eeq
where we defined the quantities $\mathcal{P}=(1+2\Omega_\gamma+2\Omega_\nu)$ and $\mathcal{X}=\left( -2 (1-\Omega_\chi) + (1-3\omega_\varphi) \Omega_\varphi / 4 \right)$. This vector differs considerably from the ones found in previous studies, which is due not only to the presence of two coupled scalar fields, but also to the improved treatment of the tight coupling approximation, which changes the subleading order contributions to the perturbation matrix $A(x)$.

Finally, the neutrino/scalar isocurvature mode is characterized by which $S_{\gamma,b}=0$, as this is the only relative entropy perturbation not involving neutrinos or scalar fields. Enforcing this necessarily implies $S_{\varphi,\chi}=\Gamma_\varphi = \Gamma_\chi = 0$, with all other relative entropy perturbations non-vanishing. The corresponding eigenvector reads at leading order: 
\beq
\begin{pmatrix}
\Delta_\nu \\
V_\nu \\
\Delta_\gamma \\
\Delta_b \\
V_{\gamma b} \\
\Delta_\varphi \\
V_\varphi \\
\Delta_\chi \\
V_\chi \\
\Pi_\nu
\end{pmatrix}_{\rm neut. iso.}
= \mathcal{C}
\begin{pmatrix}
\mathcal{R} \\ 15/4 \\ -\Omega_\nu / \Omega_\gamma \, \mathcal{R} \\ 3 \Omega_\nu/4\Omega_\gamma \, \mathcal{R} \\-\Omega_\nu/4\Omega_\gamma \,(\mathcal{R} + 4 \Omega_\gamma) \\ 0 \\ -\Omega_\nu \\0\\-\Omega_\nu \\ 3 
\end{pmatrix} \, ,
\eeq

Note that due to the convenient choice of perturbation variables, the dependence on the coupling only appears at subleading order in the baryon isocurvature mode through $\omega_\varphi$. At leading order, the nullspace of the leading order perturbation matrix $A_0$ is completely independent of $r_\varphi$ or $q_\varphi$.

\section{Coupled exponential potentials}
\label{sec:coupledExponentials}

In this section we investigate one specific example of two coupled canonical scalar fields given by
\beq
\label{expPot}
V(\varphi,\chi) = M^4 \left[ {\rm e}^{-\alpha \varphi/M} + \mu \, {\rm e}^{-2\beta \varphi/M} {\rm e}^{\lambda \chi / M} \right] \, .
\eeq
We chose this particular shape for two reasons. First, it arises in the cosmon-bolon model for quintessence and dark matter \cite{Beyer:2010mt}, which gets investigated more deeply in an accompanying paper \cite{CB_LinPers}. Furthermore, as we will show in the next section, this simple potential is already the most generic case, in the sense that all exact scaling solutions existing for other potentials are already present for this exponential form.

One can easily check that this potential fulfills equation (\ref{lnVConstraint}) with $c_1=3(1+\omega_{\rm eff})/\alpha$, $c_2=4(2\beta-\alpha)/\alpha \lambda$ and 
\beq
f(\xi_2) = {\rm ln} \left[ M^4 \left( 1+ \mu {\rm e}^{\lambda \xi_2/M} \right) \right] \, .
\eeq
For the subsequent analysis we split the potential as follows
\beq
\label{GenericTwoScalarPotential}
V(\varphi,\chi) = V_1(\varphi) +  V_2(\varphi,\chi) \,
\eeq
with 
\beq
V_2(\varphi,\chi) =M^4 \mu \,  {\rm e}^{-2 \beta \varphi/M} {\rm e}^{\lambda \chi/M} \, ,
\eeq
and assign energy- and pressure-densities accordingly:
\begin{align}
\label{varphiDef}
\rho_{\varphi}= X_{\varphi} + V_1\, , \quad p_{\varphi} = X_{\varphi} - V_1 \, , \\
\label{chiDef}
\rho_{\chi}= X_{\chi} + V_2\, , \quad p_{\chi} = X_{\chi} - V_2 \, . 
\end{align}

\begin{center}
\begin{table*}[t]{
\renewcommand{\arraystretch}{2}
\begin{tabular}{c | c c c c c c}
\hline
{\rm Point} & x & y & u & v & z & $\omega_{\rm eff}$ \\
\hline
K$_1$ & $\sqrt{6}/\alpha$ & $0$ & $\pm \left(1- \frac{6}{\alpha^2} \right)^{1/2}$ & $0$ & $0$ & $1$\\
\hline
K$_2$ & $\frac{2 \sqrt{6} \beta \pm \lambda  \sqrt{f-6}}{f}$ & $0$ & $\frac{\pm 2 \beta  \sqrt{f-6} - \sqrt{6} \lambda }{f}$ & $0$ & $0$ & $1$\\
\hline
S$_1$ & $\sqrt{\frac{2}{3}} \beta$ & $0$ & $- \lambda/\sqrt{6}$ & $\left(1-\frac{f}{6}\right)^{1/2} $ & $0$ & $-1+f/3$ \\
\hline  
S$_2$ & $\alpha/\sqrt{6}$ & $\left( 1-\frac{\alpha^2}{6} \right)^{1/2}$ & $0$ & $0$ & $0$ & $-1+ \alpha^2/3$\\
\hline
S$_3$ & $\frac{\alpha  \lambda ^2}{\sqrt{6} g}$ & $\left({\frac{(f - 2\beta \alpha)(6g-\alpha^2 \lambda^2)}{6g^2}}\right)^{1/2}$ & $\frac{\alpha (2 \beta -\alpha) \lambda }{\sqrt{6} g}$ & $\left(\frac{\alpha (\alpha - 2\beta)(6g - \alpha^2 \lambda^2)}{6g^2}\right)^{1/2}$ & $0$ & $-1+\frac{\alpha^2 \lambda^2}{3g}$ \\
\hline
R$_1$ & $0$ & $0$ & $0$ & $0$ & $1$ & $1/3$ \\
\hline
R$_2$ & $\sqrt{\frac{2}{3}} \frac{4 \beta}{f}$ & $0$ & $- \sqrt{\frac{2}{3}} \frac{2\lambda}{f}$ & $\left( \frac{4}{3f}\right)^{1/2}$ & $1 - \frac{4}{f}$ & $1/3$\\
 \hline
R$_3$ & $\sqrt{\frac{2}{3}}\frac{2}{\alpha}$ & $\frac{2}{\sqrt{3} \alpha}$ & $0$ & $0$ & $1-\frac{4}{\alpha^2}$ & $1/3$\\
\hline
R$_4$ & $\sqrt{\frac{2}{3}}\frac{2}{\alpha }$ & $\frac{2 \sqrt{f-2 \alpha  \beta}}{\sqrt{3} \alpha  \lambda }$ & $-\sqrt{\frac{2}{3}} \frac{2 (\alpha -2 \beta )}{\alpha  \lambda }$ & $\left( \frac{4 (\alpha -2 \beta)}{3 \alpha \lambda^2} \right)^{1/2}$ & $1 - \frac{4 g}{\alpha ^2 \lambda ^2}$ & $1/3$\\
\hline
\end{tabular}
}
\caption{Fixed points for coupled exponential potentials. Here $f(\beta,\lambda) = 4 \beta^2 + \lambda^2$, $g(\alpha, \beta, \lambda) = (\alpha-2\beta)^2 + \lambda^2$
and the sign change for the point $K_2$ should be taken simultaneously in $x$ and $y$.}
\label{fixedPoints}
\end{table*}
\end{center}

\subsection{Exact scaling solutions}

We start by finding all exact scaling solutions for the case of two coupled canonical scalar fields $\varphi$ and $\chi$ in the presence of a radiation fluid, which is the relevant case for studying physics in the early universe. 
The equations of motion governing this system can be directly read off from equations (\ref{scalarFEQ}) and (\ref{matterEEC}). As is common when trying to find exact scaling solutions, we now employ the dynamical systems approach \cite{Copeland:1997et,Gumjudpai:2005ry} and introduce the following new variables:
\begin{align}
 x &= \frac{\varphi'}{\sqrt{6} M h} \, ,  \quad y  = \frac{a \sqrt{V_1}}{\sqrt{3}Mh} \, , \\
 u &= \frac{\chi'}{\sqrt{6}Mh} \, , \quad v=\frac{a \sqrt{V_2}}{\sqrt{3}Mh} \, ,\\
 z &= \frac{a \sqrt{\rho_r}}{\sqrt{3}Mh} \, ,
\end{align}
where these variables are subject to the constraint
\beq
z^2=1-x^2-y^2-u^2-v^2 \, ,
\eeq
since are assuming a flat universe. The equations of motion can now be rewritten as
\begin{align}
\label{xEQ}
\frac{\d x}{\d N} = & 2 x \left(1-u^2-v^2-x^2-y^2\right)+3 u^2 x \nonumber \\
& +\sqrt{6} \beta  v^2+3 x^3-3 x+\frac{1}{2} \sqrt{6} \alpha  y^2 \, , \\
\frac{\d y}{\d N} =& 2 y \left(1-u^2-v^2-x^2-y^2\right)+3 u^2 y \nonumber \\
&+3 x^2 y-\frac{1}{2} \sqrt{6} \alpha  x y \, , 
\end{align}
\begin{align}
\frac{\d u}{\d N} =& 3 u^3+2 u \left(1-u^2-v^2-x^2-y^2\right)+3 u x^2 \nonumber \\
& -3 u-\frac{1}{2} \sqrt{6} \lambda  v^2 \, , \\
\label{vEQ}
\frac{\d v}{\d N} =& 2 v \left(-u^2-v^2-x^2-y^2+1\right)+3 u^2 v \nonumber \\
&+\frac{1}{2} \sqrt{6} \lambda  u v+3 v x^2-\sqrt{6} \beta  v x \, , 
\end{align}
and a fixed point is characterized by $\d x / \d N = \d y / \d N = \d u / \d N = \d v / \d N = 0$. When solving the resulting algebraic equations we can reduce the number of fixed points by eliminating some redundancies. First we assume that $\alpha>0$ and $\lambda>0$, both can be easily achieved by a suitable sign-change in the fields $\varphi$ and $\chi$. Furthermore we have $y>0$ and $v>0$ by definition. Restricting ourselves to these ranges, the complete set of all fixed points is given in Table \ref{fixedPoints}.

These fixed points can be split up into the purely scalar field dominated fixed points $K_1$ to $S_3$ and radiation-like fixed points $R_1$ to $R_4$. We expect the kinetically dominated fixed points $K_1$ and $K_2$ to always exist, irrespective of the exact shape of the scalar potential, as long as it vanishes asymptotically. Since the kinetic energies are non-zero whereas the potential energies vanish, this requires an 'extended region of a zero potential' where the fields can roll freely, a scenario reached asymptotically in the limit $\varphi \rightarrow \infty$ for our potential. Furthermore the point $R1$ corresponds to both scalar fields sitting at a root of the potential, in the case of $\varphi$ at infinity. This point is present even more generically, all that is required of the potential is a (possibly asymptotic) root. However, these points are always unstable, as we will see below.

The remaining points $S_1$ to $S_3$ and $R_2$ to $R_4$ are the ones only existing for potentials of the form derived in section \ref{sec:GenericScalingSolutions} and among these, only the ones exhibiting a radiation-like expansion, i.e. $R_2$ to $R_4$, can result in a realistic early cosmology.

\begin{table}[b]
\renewcommand{\arraystretch}{1.5}
\setlength{\tabcolsep}{7pt}
\begin{tabular}{c | c c }
\hline
{\rm Point} & Existence & Stability \\
\hline
K$_1$ & $\alpha^2 \geq 6$ & unstable $\forall \alpha, \beta, \lambda$\\
\hline
K$_2$ & $f(\beta,\lambda) \geq 6$& unstable $\forall \alpha, \beta, \lambda$\\
\hline
S$_1$ & $f(\beta,\lambda) \leq 6$ & $f(\beta,\lambda) <  {\rm min}(4, 2 \alpha \beta)$\\
\hline
S$_2$ &  $\alpha^2 \leq 6$ & $\alpha < {\rm min}(2,2\beta)$\\
\hline
S$_3$ & $f(\beta,\lambda), \alpha^2 \geq 2 \alpha \beta$,  & $f(\beta,\lambda), \alpha^2 > 2 \alpha \beta$\\
 & $6 g \geq \alpha^2 \lambda^2 $& $4 g >\alpha^2 \lambda^2$\\
\hline
R$_1$ & always & unstable $\forall \alpha, \beta, \lambda$ \\
\hline
R$_2$ & $f(\beta,\lambda) \geq 4 $ & $2 \alpha \beta > f(\beta,\lambda)> 4 $ \\
\hline
R$_3$ & $\alpha^2 \geq 4$ & $2<\alpha<2 \beta$ \\
\hline
R$_4$ & $f(\beta,\lambda), \alpha^2 \geq 2 \alpha \beta$,  & $4 g < \alpha^2 \lambda^2$, \\
 & $4 g \leq \alpha^2 \lambda^2$ & $\alpha \lambda^2 > 2 \beta g $\\
 \hline
\end{tabular}
\caption{Conditions for existence and stability of the fixed points.}
\label{stabilityTable}
\end{table}
The stability of the fixed points can be analyzed by investigating the Jacobian matrix for $x,y,u$ and $v$ for each fixed point. The typical classification distinguishes between nodes, spirals and saddle-points as well as stable and unstable points \cite{Copeland:1997et,Gumjudpai:2005ry}. Here we are not interested in the precise details and simply call points for which the Jacobian matrix has only positive (or zero) eigenvalues unstable, and points for which all eigenvalues are negative stable. The conditions for existence and stability of the fixed points are given in Table \ref{stabilityTable}. 

\begin{figure}[t]
	\centering
  \includegraphics[width=1.0\linewidth]{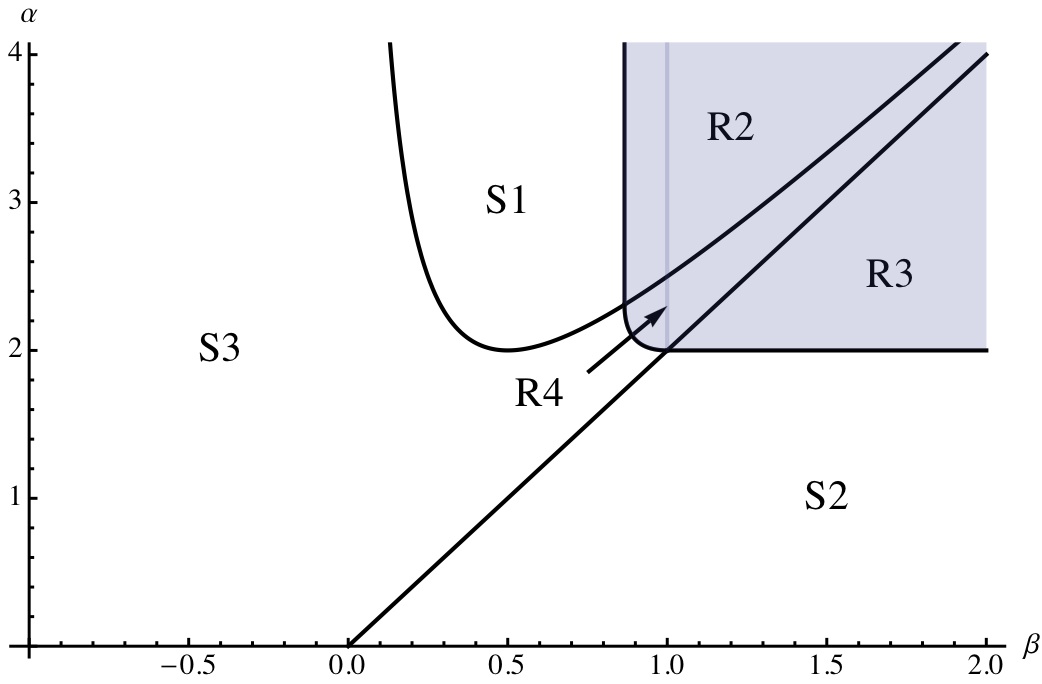}
	\caption{Parameter space of stable fixed points for the coupled exponential potential with $\lambda=1$, the structure remains valid for all $\lambda<2$. The shaded region shows the points which allow for a realistic early universe cosmology, i.e. a radiation-like expansion.}
	\label{fig:paramSpace1}
\end{figure}

\begin{figure}[t]
	\centering
  \includegraphics[width=1.0\linewidth]{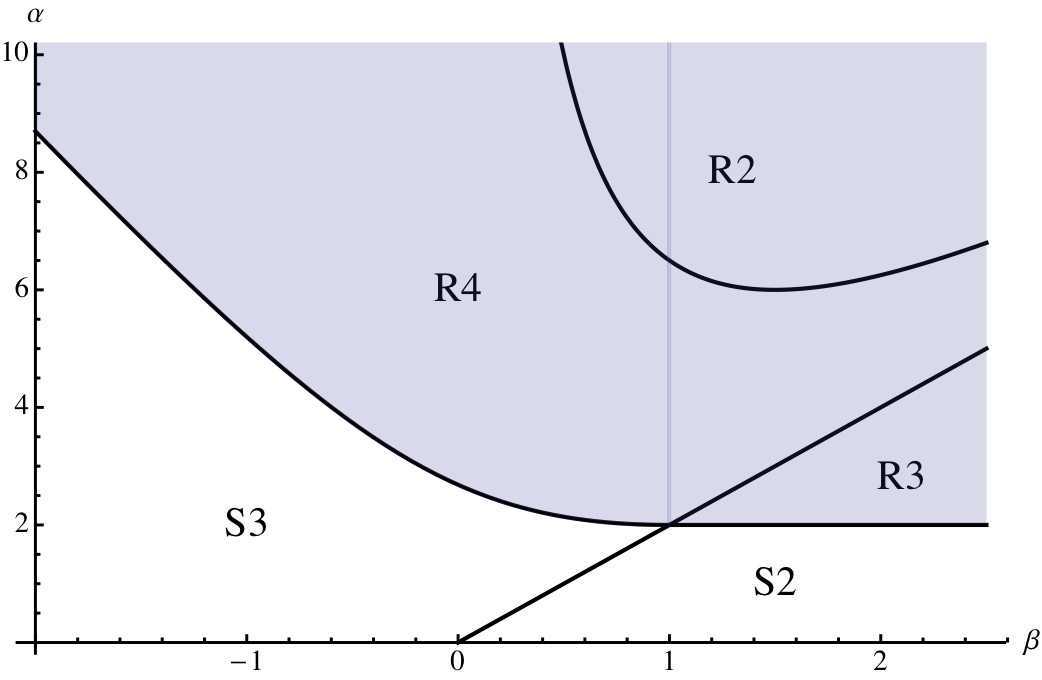}
	\caption{Parameter space of stable fixed points for the exponential coupling potential with $\lambda=3$, the structure remains valid for all $\lambda>2$. The shaded region shows the points which allow for a realistic early universe cosmology, i.e. a radiation-like expansion.}
	\label{fig:paramSpace2}
\end{figure}

As is common in scenarios such as ours, the stable fixed points split up the parameter space into disjunct sections. For a given set of parameters $\alpha, \lambda$ and $\beta$ we therefore have a unique attractive scaling solution towards which the cosmological evolution will adjust itself relatively quickly in the early universe. To avoid this fixed point for the prolonged period of radiation domination, one would necessarily have to start with initial conditions very far away from the fixed point configuration. The splitting of the parameter space can be seen in Figures \ref{fig:paramSpace1} and \ref{fig:paramSpace2}.

\subsection{Dominant perturbation modes}
The dominant perturbation modes in this potential can now be obtained in the same fashion as the general results in section \ref{sec:EarlyUniversePerturbations}, but this time with the simplifications
\beq
r_\varphi = \frac{6 \Omega_\varphi (3 \omega_\varphi -1)}{2 \Omega_\chi + \Omega_\varphi \left(3 \omega_\varphi - 1 \right)}
\eeq
and
\beq
\beta=\frac{\sqrt{3} \Omega_\varphi (3 \omega_\varphi - 1)}{\sqrt{(1+\omega_\varphi) \Omega_\varphi} \left( 2 \Omega_\chi + \Omega_\varphi (3 \omega_\varphi - 1) \right)} \, .
\eeq
The generic results for the eigenvalues do of course still hold, but as is shown in Figures \ref{fig:genericEV5_1} and \ref{fig:genericEV5_2}, the particular shape of the potential leads to a form of the coupling term $r_\varphi$ that excludes the region of strongly growing eigenvalues. This holds for all stable fixed points, with the exception of $R3$, where $\Omega_\chi = 0$ and which therefore requires some additional treatment. However, a vanishing $\Omega_\chi$ clearly requires $\omega_\varphi=1/3$ by virtue of equation (\ref{omegaScalarEQ}), effectively reducing the scenario to an uncoupled one, where all eigenvalues are non-positive.

Thus for the coupled exponential potential there is a huge region of the parameter space which is not only viable as an early cosmology at the background level, but for which no strongly growing perturbation modes exist and the adiabatic mode is therefore stable. With the results presented here one can therefore easily write down the early universe background evolution for any set of parameters $\alpha,\beta$ and $\lambda$ and also the dominant perturbation modes, given by the nullspace of $A_0$, with the density parameters replaced by the ones given in Table \ref{fixedPoints}.

\section{Non exponential models}
\label{app:otherModels}
We now move on to provide short arguments why all exact scaling solutions for coupled two scalar field cosmologies are effectively the ones found for the coupled exponential potential in section \ref{sec:coupledExponentials}. We start by recalling the generic shape of the common scalar potential required for the existence of scaling solutions given in equation (\ref{lnVConstraint}) together with the splitting (\ref{potSplitMain}), which implies via equation (\ref{ftog}) that the common potential is given by
\beq
\label{genericV2}
V_2(\varphi,\chi) = \mu_2 \, g(\xi) \, {\rm e}^{-\alpha \varphi/M} \, ,
\eeq 
where $\xi = \chi - \sigma \varphi$ with $\sigma = c_2/c_1$ and $\alpha = 3 (1+\omega_m) M /c_1$.
Our aim is to rewrite the field equations in the same form used in section \ref{sec:coupledExponentials}, in order to recover a set of algebraic equations whose fixed points determine the existing scaling solutions. For non-exponential potentials we can of course not express the potential derivatives in terms of the potential, and we therefore need a new set of variables, defined by
\beq
s_n \equiv \frac{g^{(n)}(\xi) M^n}{g(\xi)} \quad {\rm for} \quad n > 1 \, .
\eeq
The evolution equations for these variables are very simple and read
\beq
\label{dsn}
\frac{\d s_n}{\d N} = \left(s_{n+1} - s_1 s_n\right) \sqrt{6} \left(u - \sigma x  \right) \, .
\eeq
The field equations can now be rewritten as in section \ref{sec:coupledExponentials}. They read:
\begin{align}
\label{xEQG}
\frac{\d x}{\d N} = & -x + x^3 -2 x y^2 + x u^2 -2 x v^2 + \frac{\sqrt{6}}{2} \alpha y^2 
\nonumber \\
&+ \frac{\sqrt{6}}{2} (\alpha + \sigma s_1) v^2 \, , \\
\frac{\d y}{\d N} =& \left[ - \frac{\sqrt{6}}{2} \alpha x + 2 + x^2 - 2 y^2 + u^2 - 2 v^2 \right] y \, , \\
\frac{\d u}{\d N} =& -u - \frac{\sqrt{6}}{2} s_1 v^2 + x^2 u - 2 u y^2 + u^3 - 2 u v^2 \, , \\
\label{vEQG}
\frac{\d v}{\d N} =& \left[ 2 + x^2 - 2 y^2 + u^2 - 2 v^2 + \frac{\sqrt{6}}{2} s_1 u \right] v \nonumber \\
& - \frac{\sqrt{6}}{2} (\alpha + \sigma s_1) x v \, . 
\end{align}
An algebraic solution for the scaling solutions can now be found by demanding $\d x / \d N = \d y / \d N = \d u / \d N = \d v / \d N = \d s_n / \d N = 0$. Luckily, the only $s_n$-term relevant for the $x,y,u$ and $v$-equation is $s_1$, which has to be a constant. Thus, equations (\ref{xEQG}) to (\ref{vEQG}) give exactly the same system of algebraic equations as equations (\ref{xEQ}) to (\ref{vEQ}), with the replacements
\beq
\label{s1Sub}
s_1 = \lambda \quad {\rm and} \quad \sigma \lambda = -\alpha + 2 \beta \, .
\eeq

Let us quickly recap what this tells us. Any exact scaling solution for two coupled canonical scalar fields is effectively given by one of the points found above for the exponential cross-coupling potential. One should be careful not to take this conclusion too far. We emphasize that the conditions that all $s_n$ have to be static are not necessarily met by all the points above for all potentials of the shape given in equation (\ref{genericV2}). However, equations (\ref{dsn}) and (\ref{s1Sub}) tell us that all points for which $u = x (2 \beta - \alpha)/\lambda$, i.e. $S3$ and $R4$, do exist for all such models. Furthermore, as was discussed in section \ref{sec:coupledExponentials}, the points $K1,K2$ exist if $V_2$ vanishes asymptotically and $R1$ does, if $V_2$ exhibits a real root. The stability of scaling solutions may also be very different for different potentials. While the stability-analysis in the $x,y,u$ and $v$-directions proceeds precisely as in section \ref{sec:coupledExponentials}, the stability of the $s_n$ provides new constraints, which depend on the functional form of the potential.

When it comes to perturbations around the possible fixed points, in particular the question of the stability of the adiabatic perturbation mode, we can draw no conclusions in general. The reason for this is, as we have seen above, a second order potential derivative, denoted by $r_\varphi$ enters the equations, which does not appear at the background level. In the language used here, this corresponds to knowledge about the term $s_2$, which is not relevant for finding the scaling solutions.

To sum up: We have shown that all scaling solutions for two coupled canonical scalar fields are included in the set of fixed points found for the exponential potentials in section \ref{sec:coupledExponentials}, and that two of these points exist for all potentials of the allowed shape. However, the stability conditions for the fixed points can change for different potentials, and so can the growth of linear perturbations modes around them.

\subsection{'Power law couplings'}
As an example of a model involving strongly growing perturbation modes, we now investigate potentials with a power law coupling. By this we mean models for which $g(\xi) = \lambda \xi^m / M^m$, where obviously $m$ has to be even in order for the potential to be bound from below. In this case the system of algebraic equations is closed after adding the equation for $s_1$, since
\beq
s_n = \frac{m!}{(m-n)! \, m^n} s_1^n \quad \text{ for $n \leq m$} \, ,
\eeq
and of course $s_n = 0$ for $n \geq m$. In this case the evolution equations for $s_1$ reads
\beq
\frac{\d s_1}{\d N} = - \frac{\sqrt{6}}{m} s_1 (s_1 u - (2 \beta - \alpha) x) \, ,
\eeq
where we already used $\beta$ as defined in equation (\ref{s1Sub}). As we have just seen, the point $R4$ exists for all choices of $g$, and a stability analysis reveals that stability for the power-law coupling requires
\begin{align}
\label{pcR4Stability1}
s_1^2 \alpha^2 > 4 g(\alpha,\beta,s_1) \, , \quad 8 \beta + (m-4) \alpha > 0 \, , \\
\label{pcR4Stability2}
(\alpha - 2  \beta) \left( (m-2)s_1 - 2 m \beta (\alpha - 2 \beta) \right) >0 \, .
\end{align}
As we have restricted ourselves to $\alpha > 0$, it is immediately clear that stability is not possible for $m=2$, but already for $m=4$, the parameter range of stability is quite big, but requires a positive coupling.

In order to see how perturbations around such a scaling solutions behave, we need to calculate $r_\varphi$. A relatively straightforward calculation shows that
\beq
r_\varphi^{pc} = r_\varphi^{exp} \left( 1 - \frac{(\alpha - 2 \beta)^2}{4 \beta^2 m} \right) \, .
\eeq
Here we used the superscripts $pc$ for the 'power law coupling' models investigated in this section, and $exp$ for the exponential coupling model investigated in section \ref{sec:coupledExponentials}. Clearly  a power law coupling lowers the $r_\varphi(\omega_\varphi)$ for positive values of $r_\varphi$ curve compared to the exponential coupling model, which pushes it into the regime where strongly growing modes exist. We have visualized this in Figures \ref{fig:genericEV5_1} and \ref{fig:genericEV5_2} by a dashed red line, obtained for $m=4, \alpha=4$ and $\beta=1$. This lies well within the regime of stability for the fixed point, as the relations (\ref{pcR4Stability1}) and (\ref{pcR4Stability2}) are satisfied for this choice of parameters.

We have thus shown that it is possible to construct coupled scalar field models where stable scaling solutions with strongly growing perturbation modes exist, i.e. the adiabatic mode becomes unstable.

\section{Conclusion}
\label{sec:Conclusion}
In this paper we have analyzed exact scaling solutions in cosmologies containing multiple canonical scalar fields coupled through their potential at both the background level and at the level of linear perturbations. At the background level we have formulated the restrictions the demand that realistic scaling solutions should exist puts on the shape of the common potential and analyzed the existing scaling solutions for one specific choice of potential, and exponential coupling, in detail. As it turns out, the resulting set of scaling solutions is complete in the sense that all possible alternative potentials can not add new solutions to the list. However, the behavior of the scaling solutions concerning stability and growth of perturbations around such solutions can be different for alternative potentials.

At the level of linear perturbations we have analyzed the growth behavior of perturbations around scaling solutions in some detail. In particular we have shown that a strong coupling can dramatically change the growth of perturbation modes. For the specific case of an exponentially coupled potential however we have seen that there are no \textit{fast growing} modes, i.e. no modes growing faster than the adiabatic mode demanded to be dominant by current observations. In this sense, the adiabatic mode is stable in this model, any initially small admixture of non-adiabatic modes will remain small. This is not a trivial fact, as we have demonstrated by constructing a scenario where a fast growing perturbation mode around a stable scaling solution exists. In this case one would have to find a mechanism to suppress the initial contribution of the fast growing isocurvature modes very strongly in order to remain in agreement with observations.

The analysis presented here is of particular relevance for coupled scalar field models of dark matter, for which it is conceivable that they follow such a scaling solution during the early universe. In an accompanying paper we have investigated such models further \cite{CB_LinPers}, building on the results presented here.


\acknowledgments{This work is supported by the grant ERC-AdG-290623.}

\appendix

\section{Gauge invariant perturbations}
\label{app:LinearPerturbations}

When studying linear perturbations in a cosmological setting, the number of different gauges and conventions for perturbative quantities one can employ is quite large. In order to avoid confusion, we will use this appendix to set up the notation we are using throughout this paper. We will work within a manifestly gauge-invariant approach and use the first part of this section to specify our definitions, setting up Einstein's equations and the equations of energy- and momentum-conservation for a generic cosmology containing several non-minimally coupled fluids. In a second part we will turn to the perturbations for two canonical scalar fields with a common potential and show how they can be mapped onto the fluid description.

\subsection{Perturbative quantities}

\subsubsection{Metric perturbations}

The most generic linearly perturbed metric around a FLRW-background contains scalar, tensor and vector modes (see e.g. refs. \cite{Mukhanov:1990me,Kodama:1985bj}). Since we are investigating a theory of scalar fields, we will restrict our attention to the scalar perturbations. With this constraint the line element of the linearly perturbed FLRW-metric reads
\begin{align}
\label{PerturbedMetric}
ds^2 = a(\eta)^2 \left\{ \right. &- (1+2\phi) d\eta^2 + 2 G,_id\eta dx^i  \nonumber \\
& \left. + \left[ (1-2\psi) \delta_{ij} + F_{,ij}\right]dx^i dx^j \right\} \, ,
\end{align}
where $\eta$ denotes conformal time, defined by $\eta(t) \equiv \int_{t_0}^{t} \frac{dt'}{a(t')}$ with $t_0$ being some suitable initial cosmic time. Under a gauge-transformation $\eta \rightarrow \eta+\xi^0 (\eta,\vec{x}) \, , \; x^i \rightarrow x^i + \xi^{,i}(\eta,\vec{x})$ the metric perturbations transform like
\begin{align}
\phi \rightarrow \phi-h\xi^0-\xi^0{}'\, &,  \quad \psi \rightarrow \psi + h \xi^0 \, , \nonumber \\
G \rightarrow G + \xi^0 - \xi' \, &, \quad F \rightarrow F - \xi \, ,
\end{align}
where $'$ denotes a derivative with respect to conformal time. Note that we have ignored the vector component of gauge transformation acting on the spatial coordinates, since it does not affect scalar perturbations.
In the subsequent calculations we will always resort to the commonly used gauge-invariant Bardeen potentials \cite{Bardeen:1980kt}, defined as
\beq
\Phi \equiv \phi - h \sigma - \sigma'  \quad {\rm and} \quad \Psi \equiv \psi + h \sigma \, ,
\eeq
where $\sigma$ is given by $\sigma \equiv -G + F'$, i.e. $\sigma \rightarrow \sigma - \xi^0$ under the above gauge transformation.

\subsubsection{Fluid perturbations}

For any (non-minimally coupled) component of the cosmic fluid the most generic form of the linearly perturbed energy momentum tensor reads
\beq
T^{\mu}_{\alpha \, \nu} = (\rho_{\alpha} + p_{\alpha}) u_{\alpha}^\mu u_{\alpha \, \nu} + p_{\alpha} \delta^\mu_\nu + \pi^\mu_{\alpha \, \nu} \, .
\eeq
Here $\rho_{\alpha}$, $p_{\alpha}$ and $u_{\alpha}$ have the usual meanings of the total (i.e. background and perturbed) energy density, pressure density and four-velocity, respectively.  $\pi^\mu_\nu$ is the anisotropic stress tensor, which can be decomposed into trace free scalar part, a vector part and a tensor part. Again ignoring the vector and tensor contributions we end up with the following expression for the spatial components
\beq
\pi^i_{\alpha \, j} = \Pi_{\alpha},_{ij} - \frac{1}{3}\nabla^2\Pi_{\alpha} \delta_{ij} \, ,
\eeq
while all other components vanish. Writing the irrotational part (again neglecting vector modes) of the 4-velocity $u^\mu$ as
\beq
(u^\mu_{\alpha}) = \frac{1}{a}\left\{ (1-\phi)\, , \; v_{\alpha},^i \right\} \, ,
\eeq
we obtain the scalar part of the stress energy tensor to linear order:
\begin{align}
T^0_{\alpha \, 0} &= -\bar{\rho}_{\alpha} - \delta \rho_{\alpha} \, , \\
T^0_{\alpha \, i} &= (\bar{\rho}_{\alpha}+\bar{p}_{\alpha}) (G,_i + v_{\alpha},_i) \, ,  \\
T^i_{\alpha \, 0} &= -(\bar{\rho}_{\alpha}+\bar{p}_{\alpha}) (v_{\alpha},^i) \, ,  \\
T^i_{\alpha \, j} &= (\bar{p}_{\alpha} + \delta p_{\alpha}) \delta^i_j+ \pi^i_{\alpha \, j} \, .
\end{align}
Here we have used a bar (like $\bar{\rho}$) to mark background quantities. The anisotropic stress tensor $\pi^i_{\alpha \, j}$ is already gauge-invariant, whereas scalar quantities (like $\delta \rho_{\alpha}$ or $\delta p_{\alpha}$) transform like
\beq
\delta s \rightarrow \delta s - \bar{s}' \xi^0 
\eeq
under a generic gauge transformation and the velocity potential obeys
\beq
v \rightarrow v + \xi' .
\eeq
We will therefore define a gauge-invariant version for scalar quantities (i.e. $\delta \rho_{\alpha}$ and $\delta p_{\alpha}$) via
\beq
\delta s^{\rm (gi)} \equiv \delta s - \bar{s}' \sigma \, 
\eeq
and also define a gauge-invariant momentum-perturbation via
\beq
\left[ (\rho + p) v \right]_{\alpha}^{({\rm gi})} \equiv (\bar{\rho}_{\alpha}+\bar{p}_{\alpha}) (v_{\alpha}+F') \, .
\eeq
These definitions are of course not unique, there is a huge number of gauge invariant metric and fluid quantities one could employ (see e.g. ref. \cite{Kodama:1985bj} for a quite comprehensive overview). We choose to work with the definitions that correspond to the standard quantities used when working in the longitudinal gauge (defined by $F'=\sigma=0$).

In the following sections the superscript ''(gi)'' will be dropped since we will be using gauge-invariant quantities exclusively. We will also stop employing the bar to denote the background quantities, because the total densities will not appear in any of the equations anymore.

\subsection{Equations of motion}

\subsubsection{Einsteins Equations}

At the background level Einstein's equations $M^2 G_{\mu \nu} = T_{\mu \nu}$
give rise to the usual Friedmann equations, which read
\begin{align}
h^2 &\equiv \left( \frac{a'}{a} \right)^2 = \frac{a^2}{3M^2} \sum_\alpha \rho_\alpha \, , \\
h' &= -\frac{a^2}{6M^2} \sum_\alpha \left( \rho_\alpha + 3 p_\alpha \right) \, .
\end{align}

At linear order around a FLRW-background one obtains the usual equations for the Fourier modes of the two Bardeen potentials:
\begin{align}
\label{PoissonEquation}
k^2 \Psi &= -\frac{a^2}{2M^2} \sum_\alpha \left( \delta \rho_\alpha - 3 h \right[(\rho + p)v\left]_\alpha \right) \, ,\\
\label{dPsiEquation}
\Psi'+h\Phi &= -\frac{a^2}{2 M^2} \sum_\alpha \left[ (\rho + p) v \right]_\alpha \, ,\\
\label{ddPsiEquation}
\Psi'' + 2h\Psi' &+h \Phi' + (2h'+h^2) \Phi = \nonumber \\
& + \frac{a^2}{2M^2} \sum_\alpha \left( \delta p_\alpha - \frac{2}{3} k^2 \Pi_\alpha \right) \, .
\end{align}
In accordance with common practice we do not explicitly mark the Fourier modes in the perturbative quantities. The two Bardeen potential are related by
\beq
\label{PhiPsiEquation}
\Phi = \Psi - a^2 \Pi_{\rm tot}/M^2 \, .
\eeq
Here $M$ denotes the reduced Planck mass.

%


\subsubsection{Energy- and momentum (non-)conservation}

At the background level the equations of energy-momentum conservation $\mathcal{D}_\mu T^{\mu \nu}$ gives rise to total energy conservation, which reads
\beq
\label{totEnergyConservation}
\rho_{\rm tot}' + 3 h(\rho_{\rm tot} + p_{\rm tot}) = 0 \, .
\eeq
However, in a multicomponent cosmology each component obeys its own energy (non-)conservation equation, which reads
\beq
\label{SingleComponentEEC}
\mathcal{D}_\mu T^{\mu \nu}_{\alpha} = Q^\nu_{\alpha} \, ,
\eeq 
where the sum of all coupling terms is of course subject to the constraint from equation (\ref{totEnergyConservation}), which gives
\beq
\sum_{\alpha} Q_{\alpha}^\nu = 0 \, .
\eeq
At the background level the assumption of spatial homogeneity and isotropy require that
\beq
\label{BGCoupling}
Q^{\mu}_{\alpha} = (-a Q_{\alpha},0,0,0) \, ,
\eeq
and we define a dimensionless version of the coupling via
\beq
\label{DimensionlessCouplingDefinition}
q_{\alpha} \equiv \frac{aQ_{\alpha}}{3h(1+\omega_{\alpha})\rho_{\alpha}} \, .
\eeq
Evaluating equation (\ref{SingleComponentEEC}) then simply gives
\beq
\rho_\alpha'+3h(1+\omega_\alpha)(1-q_\alpha) \rho_\alpha = 0 \, .
\eeq

In order to write down the generic perturbed equations of energy- and momentum-conservation, we first have to specify the generic forms for energy- and momentum transfer at the linear level. This is done by perturbing the coupling four-vector $Q$, which reads
\begin{align}
\delta Q^0_{\alpha} & =  -aQ_{\alpha}(\phi+\epsilon_{\alpha}) \, , \\
\delta Q^j_{\alpha} & =  a \left[ Q_{\alpha} (v+G)+ f_{\alpha} \right],^j \, .
\end{align}
Note that we have again dropped a possible vector contribution to the spatial part of the perturbed coupling four-vector. While $f_{\alpha}$ is already gauge-invariant, $\epsilon_{\alpha}$ transforms as $\epsilon_{\alpha} \rightarrow \epsilon_{\alpha} -\frac{Q'_{\alpha}}{Q_{\alpha}}\xi^0$. We define a gauge-invariant momentum transfer via
\beq
\tau_{\alpha} = \epsilon_{\alpha} -\frac{Q'_{\alpha}}{Q_{\alpha}} \sigma \, .
\eeq
The equation of energy conservation then reads at the linear level
\begin{align}
\label{GenericECEQ}
\delta \rho_{\alpha}' =& - 3 h (\delta \rho_{\alpha} + \delta p_{\alpha}) + k^2 \left[ (\rho + p)v \right]_{\alpha} + 3 (\rho_{\alpha} + p_{\alpha}) \Psi' \nonumber \\
&+ 3 h (\rho_{\alpha} + p_{\alpha}) q_{\alpha} \Phi + 3 h (\rho_{\alpha} + p_{\alpha}) q_{\alpha} \tau_{\alpha} \, ,
\end{align}
where we have used the dimensionless coupling defined in equation (\ref{DimensionlessCouplingDefinition}).
The equation of momentum conservation gives
\begin{align}
\label{GenericMCEQ}
&\left[(\rho + p)v \right]_\alpha'  + 4 h \left[ (\rho + p)v \right]_\alpha + (\rho_\alpha + p_\alpha) \Phi + \delta p_\alpha = \nonumber \\
& + \frac{2}{3} k^2 \Pi_\alpha + 3 h q_\alpha \frac{(\rho_\alpha + p_\alpha)}{(\rho_{\rm tot}+p_{\rm tot})} \left[ (\rho + p)v \right]_{\rm tot} + a f_\alpha \, .
\end{align}

\subsection{New variables}

\subsubsection{Generic equations}

Following ref. \cite{Malik:2001rm} we now introduce new variables for all perturbative fluid quantities, which are particularly well suited to investigate the perturbation modes in the early universe in our model. We start with a redefined density contrast and velocity potential given by
\begin{align}
\Delta_\alpha \equiv & \frac{\delta \rho_\alpha}{\rho_\alpha} + \frac{\rho_\alpha'}{h \rho_\alpha} \Psi = \frac{\delta \rho_\alpha}{\rho_\alpha} -3 (1+\omega_\alpha)(1-q_\alpha)\Psi \, , \\
V_\alpha \equiv & -\frac{h \left[ (\rho + p)v \right]_\alpha}{\rho_\alpha + p_\alpha} \, .
\end{align}
The equations of motion (\ref{GenericECEQ}) and (\ref{GenericMCEQ}) can now be rewritten in terms of these new variables. This gives:
\begin{align}
\label{newEEC}
\Delta_\alpha' =& -3 h \omega _\alpha \Gamma _\alpha - k^2 (1+\omega _\alpha) V_\alpha /h \nonumber \\
&+3 (1+\omega_\alpha)  \left[ q_\alpha'  + 3 h (1+c_{a, \alpha}^2) q_{\alpha} (q_\alpha-1)\right] \Psi \nonumber \\
& + 3 (1+\omega _\alpha) q_\alpha \left(h \Phi +\Psi '\right) +3 h (1+\omega _\alpha) q_\alpha \tau _\alpha \nonumber \\
& -3 h \left( q_\alpha (1+\omega_\alpha) +c_{a,\alpha}^2 - \omega_\alpha \right) \Delta _\alpha \, , \\
\label{newMEC}
V_\alpha' =&3 h \left( c_{a,\alpha}^2 (1-q_\alpha) -q_\alpha - \frac{1+\omega_{\rm eff}}{2} \right) V_\alpha \nonumber \\
&+\frac{h c_{a,\alpha}^2}{(1+\omega _\alpha)} \Delta _\alpha + 3 h c_{a,\alpha}^2 (1-q_\alpha) \Psi \nonumber \\
&-\frac{2}{3} \frac{k^2}{h} \frac{\tilde{\Pi} _\alpha}{\omega_\alpha (1+\omega _\alpha)}  + \frac{h \omega _\alpha}{1+\omega _\alpha} \Gamma _\alpha - \frac{h}{\rho_\alpha + p_\alpha} a f_\alpha \nonumber \\
&+ \frac{2 q_\alpha}{1 + \omega_{\rm eff}} \left(h \Phi +\Psi '\right) +h \Phi \, ,
\end{align}
where we have introduced the adiabatic sound speed and the internal entropy perturbation (see next section for more details), given by
\begin{align}
\label{SoundSpeedDefinition}
c_{a,\alpha}^2 \equiv \, & \omega_\alpha + \frac{\rho_\alpha}{\rho_\alpha'} \omega_\alpha'  = \omega_\alpha - \frac{\omega_\alpha'}{3h(1-q_\alpha)(1+\omega_\alpha)} \, , \\
\Gamma_\alpha \equiv& \frac{1}{p_\alpha} \left( \delta p_\alpha - c_{a,\alpha}^2 \delta \rho_\alpha \right) \, ,
\end{align}
as well as a dimensionless quantity for the anisotropic stress
\beq
\tilde{\Pi}_\alpha \equiv \frac{h^2 \Pi_\alpha}{p_\alpha} \, . 
\eeq

Einstein's equations (\ref{PoissonEquation}) - (\ref{PhiPsiEquation}) can of course easily be adapted as well:
\begin{align}
\label{PsiSub}
\Psi =& -\frac{3}{2} \frac{\sum_\alpha \Omega_\alpha \left( \Delta_\alpha + 3 (1+\omega_\alpha) V_\alpha \right)}{x^2 + \frac{9}{2} (1+\omega_{\rm eff})} \, , \\
\label{dPsiSub}
\Psi'/h =& -\Phi + \frac{3}{2} \sum_\alpha \Omega_\alpha (1+\omega_\alpha) V_\alpha \, , \\
\label{PhiSub}
\Phi =& \Psi - 3 \sum_\alpha \Omega_\alpha \omega_\alpha \tilde{\Pi}_\alpha = \Psi - \Omega_\nu \tilde{\Pi}_\nu\, ,
\end{align}

\subsubsection{Equations for neutrinos, photons and baryons}

At this point we can already write down the equations relevant for photons, neutrinos and baryons in the early universe, but after electron-positron annihilation. Strictly speaking, the evolution of neutrinos and photons is described in terms of a multipole-expansion of the respective phase-space distribution functions. Following ref. \cite{Doran:2003xq} we truncate the neutrino expansion after the quadrupole,which leaves an additional anisotropic stress contribution $\tilde{\Pi}_\nu$ in addition to $\Delta_\nu$ and $V_\nu$. Since scalar fields, even when coupled, do not develop any anisotropic stress and the corresponding quantity for the baryon-photon plasma can be ignored for the era under consideration, we have $\tilde{\Pi}_{\rm tot} = \tilde{\Pi}_\nu$ and the evolution of this quantity is governed by the following equation \cite{Ma:1995ey}:
\beq
\label{dPinuEquation}
 \tilde{\Pi}_\nu ' = \frac{2 h}{1+3\omega_{\rm eff}} \left( \frac{8}{5} V_\nu - 2 \tilde{\Pi}_\nu \right) \, .
\eeq
Here we closed the equation by setting all higher order moments in the Boltzmann expansion to zero.  Other ways of closing the equations are more accurate for later times, but for the very early times considered here this is accurate enough \cite{Blas:2011rf}. The corresponding equations of energy- and momentum conservation can be read off from the generic equations (\ref{newEEC}) and (\ref{newMEC}), but with the substitutions $\omega_\nu = c_{a,\nu}^2 = 1/3$ and $\Gamma_\nu = q_\nu = 0$ to get
\begin{align}
\Delta_\nu' = & - \frac{4}{3} \frac{k^2}{h} V_\nu \, , \\ 
V_\nu' = & \frac{h}{4} \Delta_\nu - \frac{(1+ 3 \omega_{\rm eff})}{2} h V_\nu + 2 h \Psi - \left( \frac{1}{6}  \frac{k^2}{h} + \Omega_\nu h \right) \tilde{\Pi}_\nu \, .
\end{align}
The situation is slightly more complicated for photons and baryons, which form a strongly coupled plasma before the era of decoupling. The full equations for a fluid description (i.e. after truncation of the photon-Boltzmann expansion after the dipole) can be obtained by using the following replacement in the generic equations:
\begin{align}
& \omega_b \Gamma_b = c_{s,b}^2 (\Delta_b + 3 \Psi) \, , \quad a f_\gamma = - a f_b = \frac{4}{3} \frac{\rho_\gamma}{h \tau_c} \left( V_\gamma - V_b \right) \, , \nonumber \\
& \omega_b = c_{a,b}^2 = \omega_\gamma \Gamma_\gamma = q_\gamma = q_b = 0 \, , \quad \omega_\gamma = c_{a,\gamma}^2 = 1/3 \, ,
\end{align}
where $\tau_c \ll h^{-1}$ is the very short Thomson interaction timescale ($\tau_c = (a n_e \sigma_\tau)^{-1}$) and $c_{s,b}^2$ is the (non-adiabatic) baryonic sound speed.
As is common practice, we employ the tight coupling approximation (TCA), which can be found in many forms in the literature (e.g. \cite{Blas:2011rf,Pitrou:2010ai}) to simplify the interaction terms. The basic feature of this approximation stems from the fact that the momentum exchange between photons and baryons involves the very small interaction timescale $\tau_c$, satisfying $h \tau_c \ll 1$ in the early universe. The TCA to any given order then expresses the equations of energy- and momentum-conservation for photons and baryons to the corresponding order in $h \tau_c$. We use only the lowest order approximation here, which yields $V_{\gamma b} \equiv V_\gamma = V_b$ and $\tilde{\Pi}_\gamma = 0$. Redoing the calculations presented in ref. \cite{Blas:2011rf} to zeroth order for our variables gives the following equations for the remaining fluid-variables:
\begin{align}
\label{dDeltagamma}
 \Delta_\gamma' = & - \frac{4}{3} \frac{k^2}{h} V_{\gamma b}  \, , \\
\Delta_b' = & -\frac{k^2}{h} V_{\gamma b} \, , \\
\label{dVgammab}
V_{\gamma b}' = & \frac{h R}{1+R} \left(\frac{1}{4} \Delta_\gamma + \Psi \right) - \frac{3}{2} h (1+\omega_{\rm eff}) V_\gamma \nonumber \\
& + \frac{h R}{1+R} V_\gamma + h \Phi + \frac{h c_{s,b}^2}{1+R} (\Delta_b + 3 \Psi) \, ,
\end{align}
where $R=4 \Omega_\gamma / 3 \Omega_b \gg 1$. Note that the last equation is different from the corresponding equation presented in ref. \cite{Doran:2003xq}, where the usual uncoupled photon equation is used. Both equations agree to leading order in an $1/R$-expansion, but in the main text we will need the corresponding expansion to subleading order, and this is the correct equation to use.

A further simplification can be made by setting the baryonic sound-speed to zero, an approximation valid in the early universe \cite{Ma:1995ey,Blas:2011rf}.

\subsection{Entropy and curvature perturbations}
\label{app:Entropy}

\subsubsection{Entropy perturbations}

The pressure perturbation $\delta p_{\alpha}$ can be decomposed into an adiabatic and a non-adiabatic pressure perturbation
\beq
\delta p_\alpha = \delta p_{\alpha,{\rm nad}} + \delta p_{\alpha, {\rm ad}} \, ,
\eeq
where
\beq
\delta p_{\alpha, {\rm ad}} = c_{a,\alpha}^2 \delta \rho_\alpha \, .
\eeq
Here $c_{a,\alpha}^2$ is the {\it adiabatic sound speed}, a background quantity given by
\beq
c_{a,\alpha}^2 = \frac{p_{\alpha}'}{\rho_{\alpha}'} = \omega_\alpha - \frac{\omega_\alpha'}{3h(1+\omega_\alpha)(1-q_\alpha)} \, ,
\eeq
which should be clearly distinguished from the {\it total sound speed}, defined as
\beq
c_{ {\rm tot}, \alpha}^2 = \frac{\delta p_\alpha}{\delta \rho_\alpha} \, ,
\eeq
which is fundamentally a perturbative quantity. The non-adiabatic sound speed is often described using the dimensionless quantity $\Gamma_\alpha$ defined via
\beq
\delta p_{\alpha,{\rm nad}} = p_\alpha \Gamma_\alpha \, .
\eeq
The quantity $\Gamma_\alpha$ is often referred to as the \textit{intrinsic} entropy perturbation. In addition to this one-component quantity, there can also be entropy perturbations between different components of the cosmic fluid. This can be derived simply by noting that
\beq
\delta p_{\rm tot} = c_{{\rm tot}, a}^2 \delta \rho_{\rm tot} + p_{\rm tot} \Gamma_{\rm tot}
\eeq
where 
\beq
c_{a,{\rm tot}}^2 = p_{\rm tot}'/\rho_{\rm tot}' = \sum_{\alpha} (1-q_\alpha) c_{a,\alpha}^2 \frac{\rho_\alpha + p_\alpha}{\rho_{\rm tot} + p_{\rm tot}}
\eeq
and
\beq
p_{\rm tot} \Gamma_{\rm tot} = \sum_\alpha p_\alpha \Gamma_\alpha + p_{\rm tot} \Gamma_{\rm rel} \, .
\eeq
An easy calculation now shows that
\beq
p_{\rm \rm tot} \Gamma_{\rm rel} = \sum_\alpha (c_{a,\alpha}^2 - c_{a,{\rm tot}}^2) \delta \rho_\alpha \, .
\eeq
The condition for an adiabatic perturbation mode is usually given by demanding that all entropy perturbations vanish. In terms of our new variables we have
\begin{align}
p_{\rm tot} \Gamma_{\rm rel} = \frac{1}{2} \sum_{\alpha,\beta} & \frac{(1-q_\alpha)(1-q_\beta)(\rho_\alpha+p_\alpha)(\rho_\beta+p_\beta)}{\rho_{\rm tot}+p_{\rm tot}} \nonumber \\
& \times \left( c_{a,\alpha}^2-c_{a,\beta}^2 \right) S_{\alpha \beta}
\end{align}
where
\beq
S_{\alpha \beta} = \frac{\Delta_\alpha}{(1+\omega_\alpha) (1-q_\alpha)} - \frac{\Delta_\beta}{(1+\omega_\beta) (1-q_\beta)} \, .
\eeq
For this reason, $S_{\alpha \beta}$ is sometimes referred to as the relative entropy perturbation between the two fluid components labelled by $\alpha$ and $\beta$. It is sufficient to demand that all internal entropy perturbations and all relative entropy perturbations vanish in order to ensure adiabatic conditions. Note that in some analyses \cite{Doran:2003xq} different definitions of the relative entropy perturbations are used, but demanding a vanishing for these alternative definitions does not ensure a vanishing relative entropy perturbation in coupled scenarios (and are not even gauge-invariant in this case), and thus they are not suitable for our analysis.

\subsubsection{Curvature perturbations}
We define the gauge-invariant total curvature perturbation as
\beq
\zeta = - \Psi - h \frac{\delta \rho_{\rm tot}}{\rho_{\rm tot}'} = \frac{1}{\rho_{\rm tot} + p_{\rm tot}} \sum_\alpha \rho_\alpha \Delta_\alpha \, . 
\eeq
Perturbation modes for which $\zeta=0$ are known as isocurvature modes. Some works employ a different definition of the curvature perturbation \cite{Mukhanov:1990me}, but all definitions agree in the superhorizon limit for a flat universe, which is what is relevant for us. Interestingly, it is easy to show that (see e.g. ref. \cite{Malik:2001rm}) on the superhorizon scales
\beq
\zeta' = -\frac{h}{\rho_{\rm tot} + p_{\rm tot}} \delta p_{\rm nad} \, ,
\eeq
i.e. the total curvature perturbation is constant for adiabatic modes.

\subsection{Canonical scalar fields with a common potential}

\subsubsection{Basic equations}

Let us move on to the case of two canonical scalar fields $\varphi$ and $\chi$ with a common potential $V(\varphi,\chi)$. Variation of the scalar part of the action 
\beq
\mathcal{S}_{({\rm sc})} = - \int \sqrt{-g} \left[ \frac{1}{2} \mathcal{D}_\mu \varphi \mathcal{D}^\mu \varphi + \frac{1}{2} \mathcal{D}_\mu \chi \mathcal{D}^\mu \chi + V(\varphi,\chi) \right]
\eeq
with respect to the scalar field yields the Klein-Gordon equations
\begin{align}
\mathcal{D}_\mu \mathcal{D}^\mu \varphi - V,_{\varphi} = 0 \, , \\
\mathcal{D}_\mu \mathcal{D}^\mu \chi - V,_{\chi} = 0 \, .
\end{align}
For the spatially homogeneous and isotropic background we can insert the FLRW-metric and drop all spatial derivatives to obtain the scalar field equations equations as given in equation (\ref{scalarFEQ}).

At the level of linear perturbations we can split the scalar fields into background parts and perturbations: $\varphi = \varphi_0 + \delta \varphi$ and $\chi = \chi_0 + \delta \chi$. In order to remain manifestly gauge-invariant we will work with the redefined field perturbations
\beq
X = \delta \varphi - \varphi' \sigma \quad {\rm and} \quad Y = \delta \chi - \chi' \sigma \, .
\eeq
Plugging this and the perturbed FLRW-metric (equation (\ref{PerturbedMetric})) into the Klein-Gordon equation yields the following (manifestly gauge invariant) linearized field equations:
\begin{align}
\label{CosmonPerKleinGordon}
X'' + 2hX' + k^2 X + a^2 V_{,\varphi \varphi} X + a^2 V_{,\varphi \chi} Y \nonumber \\
+ 2 a^2 V_{,\varphi} \Phi - \varphi' \Phi' - 3 \varphi' \Psi' & = 0 \, , \\
\label{BolonPerKleinGordon}
Y'' + 2hY' + k^2 Y + a^2 V_{,\varphi \chi} X + a^2 V_{,\chi \chi} Y \nonumber \\
+ 2 a^2 V_{,\chi} \Phi - \chi' \Phi' - 3\chi' \Psi' & = 0 \, .
\end{align}
Following section \ref{sec:EarlyUniversePerturbations} (but keeping things slightly more general) we split the potential into two parts $V(\varphi,\chi) = V_1 (\varphi,\chi) + V_2 (\varphi,\chi)$, where the exact definitions of $V_1$ and $V_2$ are in principle arbitrary. We define a corresponding splitting of the energy-momentum tensor by
\begin{align}
T_{{\varphi}}^{\mu \nu} = \mathcal{D}_\mu \varphi \mathcal{D}^\mu \varphi - g_{\mu \nu} \left[ \frac{1}{2} \mathcal{D}_\tau \varphi \mathcal{D}^\tau \varphi + V_1 (\varphi,\chi)  \right] \, , \\
T_{{\chi}}^{\mu \nu} = \mathcal{D}_\mu \chi \mathcal{D}^\mu \chi - g_{\mu \nu} \left[ \frac{1}{2} \mathcal{D}_\tau \chi \mathcal{D}^\tau \chi + V_2 (\varphi,\chi)  \right] \, .
\end{align}
An evaluation at the background level (inserting the FLRW-metric and dropping spatial derivatives) yields a perfect fluid-form of this tensor with the identifications (\ref{varphiDef}) and (\ref{chiDef}). At the linear level we again recover the fluid form, this time with the following assignments:
\begin{align}
\label{deltaRhoVarphiDef}
\delta \rho_\varphi = \frac{1}{a^2} \left(\varphi' X' \right.&\left. - \Phi \varphi'^2 + a^2 V_{1,\varphi} X + a^2 V_{1,\chi} Y \right) \, , \\ 
\delta \rho_\chi = \frac{1}{a^2} \left( \chi' Y' \right.&\left. - \Phi \chi'^2 + a^2 V_{2,\varphi} X + a^2 V_{2,\chi} Y \right) \, , \\
\delta p_\varphi = \frac{1}{a^2} \left(\varphi' X' \right.&\left. - \Phi \varphi'^2 - a^2 V_{1,\varphi} X - a^2 V_{1,\chi} Y \right) \, , \\
\delta p_\chi = \frac{1}{a^2} \left( \chi' Y' \right.&\left. - \Phi \chi'^2 - a^2 V_{2,\varphi} X - a^2 V_{2,\chi} Y \right) \, , \\
&\left[ (\rho+p)v \right]_\varphi = \frac{-1}{a^2} \varphi' X \, , \\
&\left[ (\rho + p)v \right]_\chi = \frac{-1}{a^2}\chi' Y \, .
\end{align}
In order to find the correct expressions for energy- and momentum-transfer we have to evaluate the equations of energy- and momentum conservation $\mathcal{D}_\mu T_{\alpha}^{\mu \nu} = Q_{\alpha}^\nu$. At the background level (and using the definition (\ref{DimensionlessCouplingDefinition})) we recover 
\begin{align}
q_\varphi = \frac{V_{1,\chi} \chi' - V_{2,\varphi} \varphi'}{3h(1+\omega_\varphi) \rho_\varphi} \, , \\
q_\chi = - \frac{V_{1,\chi} \chi' - V_{2,\varphi} \varphi'}{3h(1+\omega_\chi) \rho_\chi} \, .
\end{align}
At the linear level we can use the linearized field equations (\ref{CosmonPerKleinGordon}) and (\ref{BolonPerKleinGordon}) to finally obtain rather complicated expressions for the energy- and momentum transfer. Since we assume that no additional couplings are present for the scalar fields (except of course to gravity) we know that $\mathcal{D}_\mu T_{\varphi}^{\mu \nu} + \mathcal{D}_\mu T_{\chi}^{\mu \nu}  = 0$. From this it is easy to see that energy- and momentum transfer of the two fields are related by
\beq
f_\chi = - f_\varphi \quad {\rm and} \quad \tau_\chi = \tau_\varphi \, ,
\eeq
and we can restrict ourselves to quoting only the expressions for the perturbed cosmon-coupling here:
\begin{widetext}
\begin{align}
a f_\varphi =  - \frac{a^2 V_{2,\varphi}}{ \varphi'} & \left[ (\rho + p)v \right]_\varphi + \frac{a^2 V_{1,\chi}}{\chi'} \left[ (\rho+p) v \right]_\chi + \frac{2  \left(\chi' V_{1,\chi}-\varphi' V_{2,\varphi}\right)}{3 h (1+\omega_{\rm eff})}  \Phi +\frac{2\left(\chi' V_{1,\chi}-\varphi' V_{2,\varphi}\right)}{3 h^2 (1+\omega_{\rm eff})}  \Psi ' \, , \\
\label{scalarMomentumPerturbation}
\left(\varphi' V_{2,\varphi}-\chi' V_{1,\chi}\right) \tau_\varphi = & +\frac{a^2 V_{2,\varphi}}{\varphi' } \delta \rho_{\varphi} + \frac{\left(a^2 \chi' V_{1,\varphi} V_{2,\varphi} - a^2 \varphi' V_{1,\chi} V_{2,\varphi} + \varphi' \chi'^2  V_{1,\varphi \chi} - \varphi'^2 \chi' V_{2,\varphi \varphi}\right) }{ (\rho_\varphi + p_\varphi) \chi' } \left[ (\rho + p)v \right]_\varphi \nonumber \\
&-\frac{a^2 V_{1,\chi}}{\chi'} \delta \rho_\chi 
- \frac{ \left( a^2 \varphi' V_{1,\chi} V_{2,\chi} - a^2 \chi' V_{1,\chi} V_{2,\varphi} + \varphi'^2 \chi' V_{2,\varphi \chi}  - \chi'^2  \varphi' V_{1,\chi \chi} \right)}{ (\rho_\chi + p_\chi) \varphi'} \left[ (\rho +p)v \right]_\chi \, .
\end{align}
\end{widetext}
Finally, the internal entropy perturbations for the scalar fields can also be extracted from the perturbed energy-momentum tensor. They read:
\begin{align}
\omega_\varphi \Gamma_\varphi= &
 \frac{2 a^2 V_{1,\varphi}}{\rho_\varphi \varphi'} \left[ (\rho+p)v \right]_\varphi 
 + \frac{2 a^2 V_{1,\chi}}{\rho_\varphi \chi'} \left[ (\rho+p)v \right]_\chi \nonumber \\
& - \frac{2 \left(V_{1,\varphi} \varphi' + V_{1,\chi} \chi' \right)}{3 h \varphi'^2 /a^2 + V_{2,\varphi} \varphi' - V_{1,\chi} \chi'} \frac{\delta \rho_\varphi}{\rho_\varphi} \, , \\
\omega_\chi \Gamma_\chi= &
 \frac{2 a^2 V_{2,\chi}}{\rho_\chi \chi'} \left[ (\rho+p)v \right]_\chi 
 + \frac{2 a^2 V_{2,\varphi}}{\rho_\chi \varphi'} \left[ (\rho+p)v \right]_\varphi \nonumber \\
& - \frac{2 \left(V_{2,\chi} \chi' + V_{2,\varphi} \varphi' \right)}{3 h \chi'^2 /a^2 + V_{1,\chi} \chi' - V_{2,\varphi} \varphi'} \frac{\delta \rho_\chi}{\rho_\chi} \, .
\end{align}

\subsubsection{Equations with new variables}

Now we want to rewrite the equations governing the linear scalar perturbations in a way that is particularly well suited for the analysis of early universe perturbations in a scaling scenario. This means we will remove all direct dependences on the potential and the field derivatives and replace them with the equation of state, the adiabatic sound-speeds and the couplings $q_\varphi$ and $q_\chi$. 

To simplify matters we make the assumption that $V_{1,\chi}=0$. This is not limiting the general model, as the splitting of the potential is arbitrary and we can thus choose to assign the entire common part to $T_\chi^{\mu \nu}$.

To do so we employ the following relations to replace the potential derivatives:
\begin{align}
V_{1,\varphi} \varphi' &= \frac{3}{2} h (1+\omega_\varphi) (1-q_\varphi) \rho_\varphi (c_{a,\varphi}^2 -1) \, , \\
\label{dV2Varphi}
V_{2,\varphi} \varphi' &= -3 h (1+\omega_\varphi) \rho_\varphi q_\varphi \, , \\
V_{2,\chi} \chi' &= -\frac{3}{2} h (1+\omega_\chi) \rho_\chi \left[ 1-c_{a,\chi}^2 + q_\chi (1+c_{a,\chi}^2) \right] \, .
\end{align}
The second derivatives can not be so easily replaced, but we can take the time derivative of equation (\ref{dV2Varphi}) to obtain
\begin{align}
V_{2,\varphi \chi} = \frac{\varphi'}{\chi'} & \left[\frac{9}{2} \frac{h^2}{a^2} q_\varphi C_\varphi -V_{2,\varphi \varphi}  - 3 \frac{h}{a^2} q_\varphi'  \right] \, ,
\end{align}
with $C_\varphi = \left( 1 + \omega_{\rm eff} + (1-q_\varphi) (1+c_{a,\varphi}^2) \right) $.

The derivative $V_{2,\varphi \varphi}$ remains in the equations and cannot be so readily replaced without introducing another second potential derivative like $V_{2,\chi \chi}$. Using these simplifications we can simplify the perturbed couplings as follows:
\begin{align}
\label{fvarphiSub}
a f_\varphi =& 3 q_\varphi (1+ \omega_\varphi) \rho_\varphi \left( \frac{2}{3} \frac{(\Psi'/h + \Phi)}{(1+\omega_{\rm eff})} - V_\varphi \right) \, , \\
\label{tauvarphiSub}
\tau_\varphi =&\frac{1}{1+\omega_\varphi} \Delta_\varphi + \left( \frac{3}{2} (1-q_\varphi) (1-c_{a,\varphi}^2) - \frac{a^2 V_{2,\varphi \varphi}}{3 q_\varphi h^2} \right) V_\varphi \nonumber \\
&- \left( \frac{3}{2} C_\varphi -\frac{q_\varphi'}{q_\varphi h} - \frac{a^2 V_{2,\varphi \varphi}}{3 q_\varphi h^2} \right) V_\chi + 3 (1-q_\varphi) \Psi \, .
\end{align}
The internal entropy perturbations now read:
\begin{align}
\label{GammavarphiSub}
\omega_\varphi \Gamma_\varphi= &
 (1-c_{a,\varphi}^2) \left( \Delta_\varphi +3 (1+\omega_\varphi) (1-q_\varphi) \right) \Psi \nonumber \\ 
&+ 3(1+\omega_\varphi) (1-q_\varphi) (1-c_{a,\varphi}^2) V_\varphi \, , \\
\label{GammachiSub}
\omega_\chi \Gamma_\chi =&
 (1-c_{a,\chi}^2) \left( \Delta_\chi +3 (1+\omega_\chi) (1-q_\chi) \right) \Psi \nonumber \\ 
& + 3 (1+\omega_\chi) \left( 1-c_{a,\chi}^2 + q_\chi (1+c_{a,\chi}^2) \right) V_\chi \nonumber \\
& - 6 (1+\omega_\chi) q_\chi V_\varphi  \, .
\end{align}
Now the generic perturbation equations can simply be obtained by inserting these results into the generic equations (\ref{newEEC}) and (\ref{newMEC}). However, we see that the perturbed momentum transfer cannot be fully specified without making some additional assumption about the coupling to eliminate the remaining potential derivative. This also holds in the case of scalar scaling scenarios, as is discussed in the main text.

\end{document}